\documentclass[usenatbib,useAMS,usedcolumn]{mnras}

\usepackage{epsfig}
\usepackage{times,pdflscape}
\pdfoutput=1

\hypersetup{pdfauthor={H.L. Liu},
            pdftitle={The Complete Local Volume Groups Sample - I. Sample Selection and X-ray Properties of the High-Richness Subsample},
            pdfkeywords={galaxies: groups: general, X-rays: galaxies, X-rays: galaxies: clusters, galaxies: clusters: general, galaxies: clusters: intracluster medium, galaxies: active},
            bookmarksnumbered=true}
\newcommand{\NHH}{\ensuremath{N_{\mathrm{H_{2}}}}}

\newcommand{\cm}{\ensuremath{\mbox{~cm}}}

\newcommand{\pcmcu}{\ensuremath{\cm^{-3}}}

\newcommand{\thco}{$^{13}$CO}

\newcommand{\etco}{C$^{18}$O}
\newcommand{\vel}{km\,s$^{-1}$}
\newcommand{\filname}{G350.5}
\newcommand{\filAname}{G350.5-N}
\newcommand{\filBname}{G350.5-S}
\newcommand{\imcoor}{$\alpha_{2000}=17^{\mathrm{h}}18^{\mathrm{m}}13\fs84,\ \delta_{2000}=-36\degr28\arcmin21\farcs5$}
\newcommand{\her}{$Herschel$}
\newcommand{\mline}{$M_{\rm line}$}
\newcommand{\msun}{$M_{\odot}$}
\newcommand{\lsun}{$L_{\odot}$}
\newcommand{\um}{$\mu$m}
\newcommand{\cmcm}{cm$^{-2}$}
\newcommand{\egcite}{\citep[e.g.,][]}
\newcommand{\lmsun}{$M_{\odot}$~pc$^{-1}$}
\graphicspath{{./0figure/}}

\begin{document}

\title[ 
The straight and isolated G350.54$+$0.69 filament: density profile and
star formation content
] 
{ 
The straight and isolated G350.54$+$0.69 filament: density profile and
star formation content
}

\author[H.L. Liu et al.] {Hong-Li Liu\footnotemark$^{1,2,3}$, Amelia Stutz$^3$, Jing-Hua Yuan$^{4}$ \\
  $^1$ Department of Physics, The Chinese University of Hong Kong, Shatin, NT, Hong Kong SAR \\
  $^2$ Chinese Academy of Sciences, South America Center for Astrophysics, Camino El Observatorio 1515, Las Condes, Santiago, Chile \\ 
  $^3$ Departamento de Astronom\'ia, Universidad de Concepci\'on, Av. Esteban Iturra s/n, Distrito Universitario, 160-C, Chile \\  
  $^4$ National Astronomical Observatories, Chinese Academy of Sciences, 20A Datun Road, Chaoyang District, Beijing 100012, China \\ 
}

\date{Accepted xxxx; Received xxx; in original xxxx}

\pagerange{\pageref{firstpage}--\pageref{lastpage}} \pubyear{2017}

\maketitle

\label{firstpage}

\begin{abstract} 
We investigate the global properties of the straight and isolated
filamentary cloud G350.54$+$0.69 using \her\  continuum and APEX
molecular line data. 
The overall straight morphology is similar to two other well studied
nearby filaments (Musca and Taurus-B211/3) while the isolated nature
of G350.54$+$0.69 appears similar to Musca.
G350.54$+$0.69 is composed of two distinct filaments with a length $\sim5.9$\,pc
for G350.5-N ($\sim2.3$\,pc for G350.5-S), a total mass of $\sim 810$\,\msun~($\sim 110$\,\msun),
and a mean temperature of  $\sim 18.2$\,K~($\sim 17.7$\,K). 
We identify 9 dense and
gravitationally bound cores in the whole cloud G350.54$+$0.69. The separations between cores and the
line mass of the whole cloud appear to follow the predictions of the
``sausage" instability theory, which suggests that G350.54$+$0.69 could
have undergone radial collapse and fragmentation. The presence of
young protostars is consistent with this hypothesis.
The line masses of the two filaments ($\sim 120$\,\lmsun\ for \filAname, and $\sim 45$\,\lmsun\ for \filBname), 
mass-size distributions of
the dense cores, and low-mass protostars collectively suggest that
G350.54$+$0.69 is a site of ongoing low-mass star formation.
Based on the above evidence, we place G350.54$+$0.69 in an intermediate
evolutionary state between Musca and Taurus-B211/3.
We suggest that investigations into straight (and isolated) versus
those distributed inside molecular clouds may provide important clues
into filament formation and evolution.

\end{abstract}

\begin{keywords}
  ISM: individual objects: G350.5 -- ISM: clouds -- ISM: structure -- ISM: molecules -- stars: formation -- infrared: ISM
\end{keywords}

\footnotetext[1]{E-mail: hl.liu@cuhk.edu.hk or hongliliu2012@gmail.com}
%\footnotetext[2]{E-mail: hbli@phy.cuhk.edu.hk}

\section{Introduction}
\label{sec:intro}

Filamentary structures in the ISM have long been recognized \citep[e.g., Barnard 1905,][]{sch79}.  
However, their role in the process of star formation has received renewed and focused attention
recently thanks to long-wavelength data provided by the Herschel Observatory. These data demonstrate 
the ubiquity of the filaments in the cold interstellar medium \egcite{mol10,and10,and14,stu15}.
Combined with multi-wavelength information, these data reveal a close connection
between filamentary clouds and star formation.  For example, \citet{kon15} and \citet{stu15} demonstrate that most
cores are located on the filaments within clouds.  Moreover, using gas
velocity information, \citet{stu16} show that protostellar cores are also kinematically coupled to the dense filamentary
environments in Orion; that is, the protostellar cores have similar radial velocities as the gas filament as well as being located on or
very near the filament ridgelines.

On scales larger than cores, Herschel results have focused on the column density profiles of the filaments themselves.  
\citet{arz11} found that the filaments in the low mass IC5146 molecular
cloud have an approximately uniform inner width of 0.1\,pc (assuming a distance of 460\,pc; see their appendix A).  
This result was later confirmed in other nearby and low mass clouds in the Gould Belt \citep{koc15}.  A proposed 
explanation for this observation is that the common width may be attributed to the sonic scale below which
the interstellar turbulence becomes subsonic in diffuse, and non star-forming molecular gas 
\citep{pad01,fed16,and16,and17}. Another possibility is that the approximately uniform
widths may be rooted in the dissipation of magneto-hydrodynamic (MHD)
waves \citep{hen13,and16,and17}.

Interestingly, claims of a ``universal" and constant width for filaments is being challenged by analysis in other more distant (and
sometimes higher mass) clouds, most of which lie beyond the Gould Belt.  Wider widths (between 0.26\,pc and 0.34\,pc) have been
reported in both the DR\,21 ridge and Cygnus\,X \citep{hen12}. In addition, on the basis of a large sample of low and high-mass
filaments within the Planck cold clumps, \citet{juv12} find filament widths of 0.1 to 1\,pc, with a typical value of $\sim0.2-0.3$\,pc.  
This large variation may be a result of inconsistencies both in the definition of filament widths and in the methodology
adopted for measuring this parameter \egcite{juv12,smi14,pan17}.
In further tension with the results inferred in low-mass Gould Belt cloud, the Orion A Integral Shaped
Filament (ISF) has an average profile that is extremely well represented by a power-law down to 0.05\,pc, the resolution limit of
the Herschel observations (\citet{stu16}; see their Figure 5). This profile is inconsistent with a $\sim0.1$\,pc width since such a width
would require a flattening or softening of the density profile inside that radius.  \citet{stu16} propose that the differences in the
ISF density profile and properties may be rooted in two distinct, if related, physical parameters in this filament: its high mass and the
action of the magnetic field.  One thing is clear: this varied patchwork of inconsistent results are difficult to  interpret even though
\citet{and17} argues that the 0.1\,pc width may be physically meaningful.  It is,
therefore, necessary to investigate the properties, including the density profile shape, of star forming gas filaments beyond the Gould
Belt.

\begin{figure*}
\centering
\includegraphics[width=3.4 in,height=3.36 in]{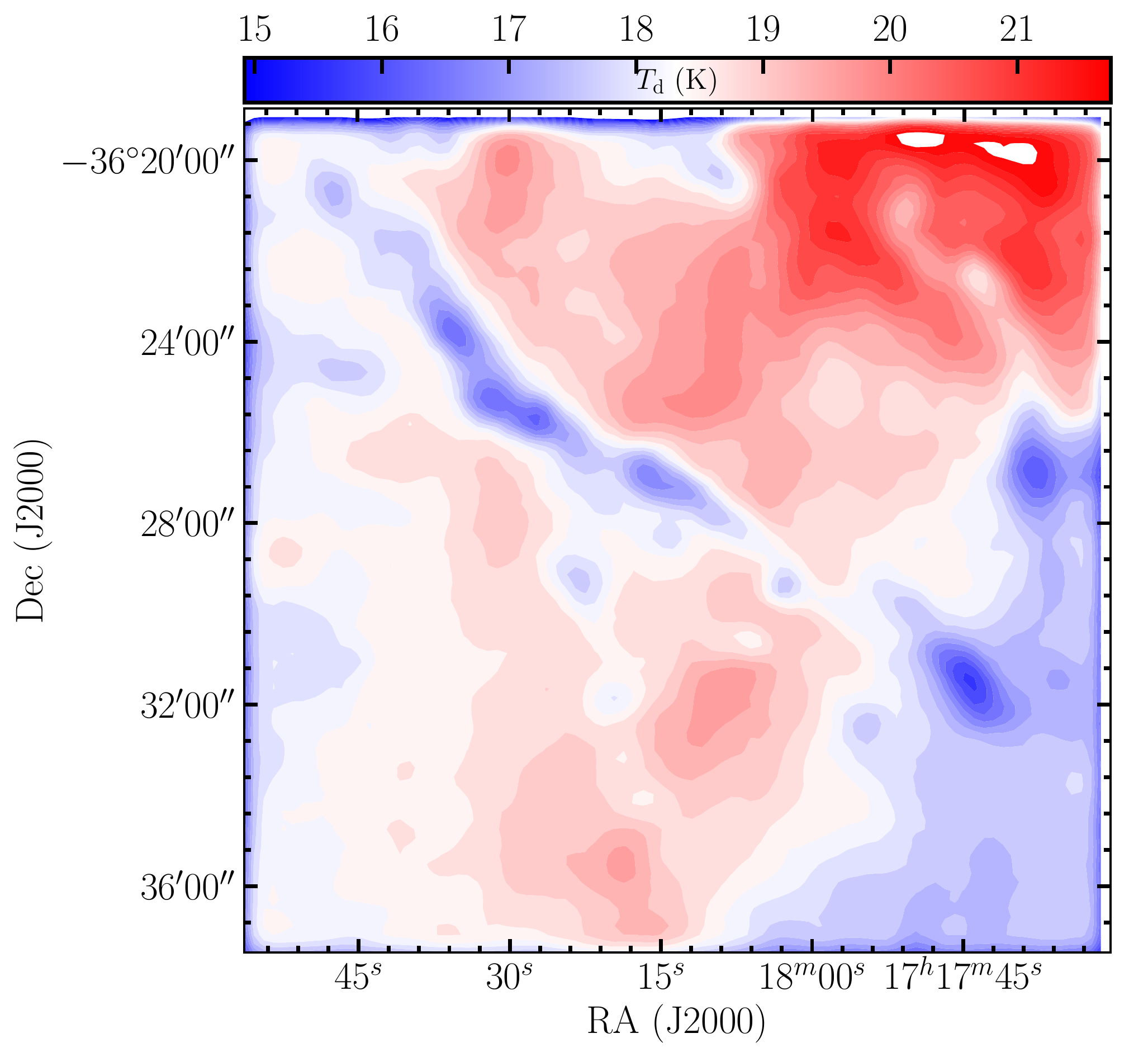}
\includegraphics[width=3.4 in]{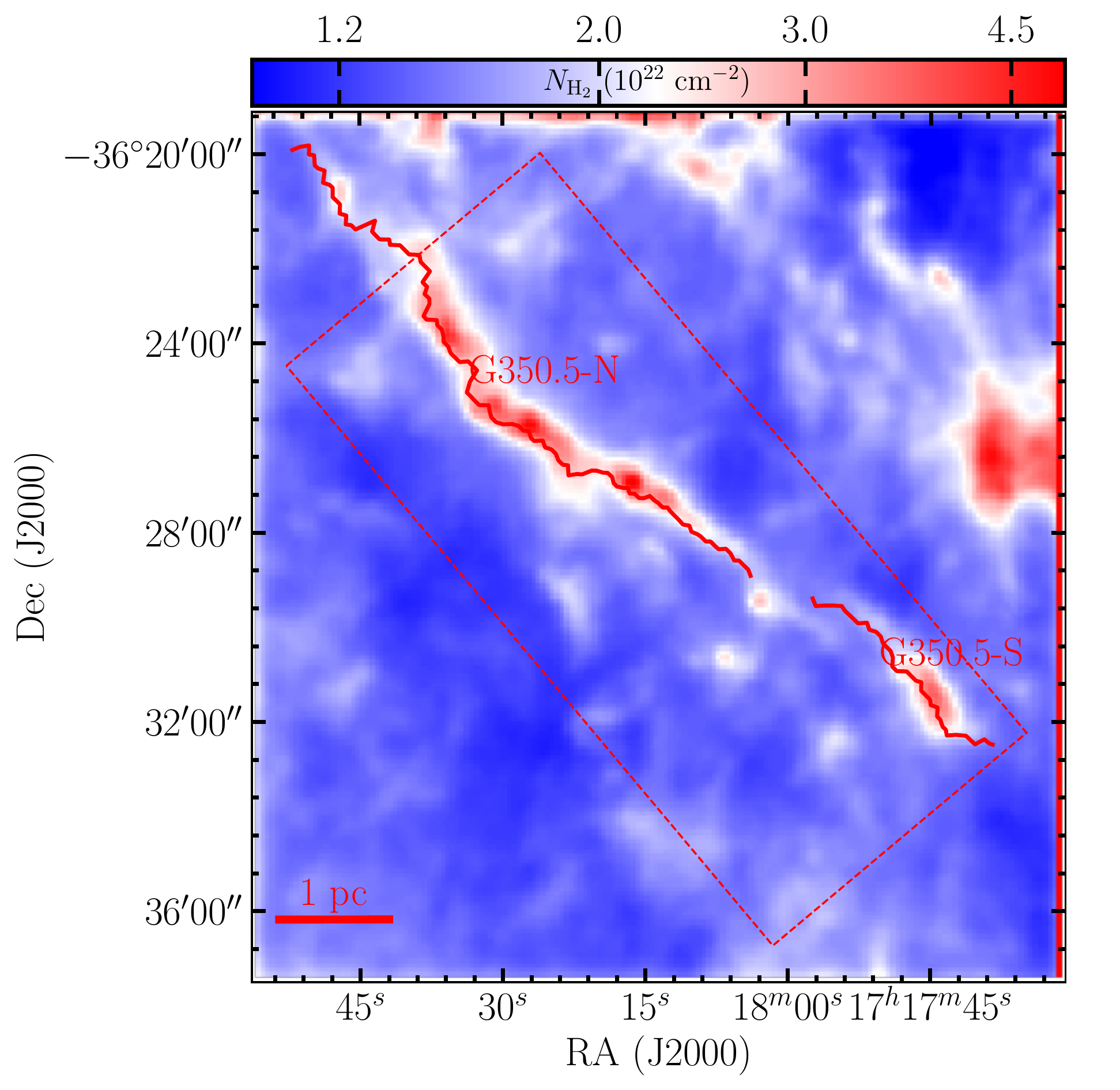}
\caption{{\it (Left):} $35\arcsec$-resolution $T_{\rm dust}$ map for the filament\,\filname. {\it (Right):}
18$\arcsec$-resolution \NHH\ column density map, where two discontinuous filaments (i.e., \filAname, and \filBname) are 
delineated with red curves. The red dashed rectangle shows the area covered by molecular line observations with APEX, 
which is rotated by 50 degrees relative to the RA direction. 
A scale bar of 1\,pc is shown on the bottom left.
}
\label{fig:filMorph}
\end{figure*}

The purpose of this paper is to investigate the evolutionary and physical state of the star forming filamentary cloud G350.54$+$0.69 (\filname\ afterwards). 
This filament is very interesting because of its straight and isolated nature (see Fig.\,\ref{fig:filMorph}), similar in morphology to 
B211/3 \citep{pal13} and in both morphology and isolation to Musca \egcite{cox16}. 
In contrast, most well-studied filaments in nearby clouds appear as curved and intertwined structures embedded in the larger molecular cloud material.   
One example includes the Orion Integral Shaped Filament (ISF), which completely dominates the cloud structure in the north of Orion A, 
has a high line-mass, and a curved wave-like morphology \citep{stu16,stu18}.  Other examples are the lower line-mass filaments embedded 
within clouds; these exhibit a "noodle soup"-like morphology, e.g, Polaris \citep{and10}, IC5146 \citep{arz11}, Pipe \citep{per12}, 
Orion\,B \citep{sch13}, and Aquila \citep{kon15}. Determining the physical nature of these two classes of filaments (i.e., 
straight and isolated versus curved and intertwined) requires detailed investigation of such straight and isolated filament, 
of which there are few examples in the literature.

 We place
particular emphasis on the filament column density profile, the kinematics of dense cores, and the star formation content of the
filaments.  G350.5 is centered at \imcoor\ and
located at $1.38\pm0.13$\,kpc (calculated using the online program Bayesian
Distance Calculator\footnote{\url{http://bessel.vlbi-astrometry.org/bayesian}} by inputing a systemic velocity of $V_{\rm lsr}=-3.9$\,\vel. 
The analysis on the column density profile can, on one hand, add to a number of previous studies of nearby filament clouds for
pinpointing which properties may give rise to the inner width of a given filament. Moreover, G350.5 has not been well-studied up till
now, and therefore represents a new laboratory within which to investigate the physical conditions in star forming filaments.
Finally, we tentatively place G350.5 in an evolutionary context through comparison to the nearby Taurus B211/3 \citep{pal13} 
and Musca \egcite{cox16} filaments, which exhibit similar morphology to G350.5.

   This paper is organized as follows: the \her\ and molecular line observations are described in Section 2, the data analysis results are presented
   in Section 3, the discussions are in Section 4, and a summary is given in Section 5.

\section{Observations and data reduction}
\label{sec:observations}

\subsection{{\it Herschel} observations}
Far-infrared data carried out with the {\it Herschel Space Observatory (HSO)} were used to 
characterize dust properties of the \filname\ filament. \filname\ was observed as part of the Hi-GAL\footnote{Hi-GAL, the {\it Herschel} infrared 
Galactic Plane Survey, is an Open Time Key project on the {\it Herschel Space Observatory (HSO)} aiming 
to map the entire Galactic Plane in five infrared bands. This survey covers a $|b|~<~1\degr$ wide strip
of the Milky Way Galactic plane in the longitude range $-60\degr~<~l~<60\degr$.} survey. The PACS \citep{pog10}
and SPIRE \citep{gri10} instruments were used to carry out the survey simultaneously at 70, 160, 250, 350, and 500\,\um\ in the parallel
photometric mode. The observations were made in two orthogonal scanning directions at a scan speed of 
60\,arcsec\,s$^{-1}$. The effective angular resolutions for those five wavelength bands are 6\farcs0, 12\farcs0, 18\farcs0,
24\farcs0, and 35\farcs0 \citep{mol16}, respectively, which correspond to 0.04, 0.08, 0.12, 0.16, and 0.23\,pc, respectively, at a distance of
1.38\,kpc. More details about the data preprocessing up to usable high-quality images can be found in \citet{tra11}.

\subsection{Molecular observations}

The molecular observations of \thco\ and \etco~(2-1) presented in this paper were made
on September 24, 2017, with the Atacama Pathfinder Experiment
(APEX) 12-m telescope \citep{gus06} at Llano de
Chajnantor (Chilean Andes). In the observations, the frontend was equipped with the APEX-1 receiver of the Swedish Heterodyne
Facility Instrument \citep[SHeFI,][]{vas08}  and the backend was 
the eXtended bandwidth Fast Fourier Transform Spectrometer (XFFTS) with an effective
spectral resolution of 114\,KHz or 0.15\,\vel\ at a tuned central frequency of 220\,GHz between
$\nu=220.39684$\,GHz (i.e., \thco~(2-1)) and $\nu=219.560358$\,GHz (i.e., \etco~(2-1)). The angular resolution at
these two frequencies is $\sim 28\arcsec$.

Mapping observations were made  using the on-the-fly mode in the two orthogonal 
scanning directions along 50\degr-rotated Right Ascension and Declination (see the dashed rectangle in Fig.\,\ref{fig:filMorph}). 
The mapping is centered on \imcoor\ with a rectangular size of $7\arcmin \times16\arcmin$. 
The observations were calibrated with the chopper-wheel technique and the output intensity
scale given by the system was $T_{\rm a}^{*}$, which represents the
antenna temperature corrected for atmospheric attenuation. That
intensity scale was further converted to the main-beam brightness
temperature by $T_{\rm mb} = T_{\rm a}^{*}/\eta_{\rm mb}$, where the main beam efficiency $\eta_{\rm mb}$ 
is 0.75. All observation data were reduced using the CLASS90 programme of the IRAM’s
GILDAS software \citep{gui00}. The reduced spectra finally present a typical rms value of 0.42\,K.

\section{Results}
\subsection{Dust temperature and column density maps}
\label{sec:dustmaps}
The dust temperature ($T_{\rm dust}$) and column density ($N_{\rm H_2}$) maps of the filament \filname\ were created using a modified blackbody
model to fit the spectral energy distribution (SED) pixel by pixel, as described in \citet{liu16,liu17}. Before the SED fitting, all Herschel images
except for the 70\,\um\ one were convolved to the resolution of the 500\,\um\ image, and then regridded to the same
pixel size as that in the 500\,\um\ band. Emission at 70\,\um\ was not considered in the SED fitting for a single temperature since 70\,\um\ radiation
may trace hotter components such as very small grains (VSGs), and warmer material heated by protostars, both of which can not be interpreted properly by 
a single temperature.
In the SED fitting, a dust opacity law of $\kappa_{\nu}=0.1 \times (\nu/1\,{\rm THz})^{\beta}$ was assumed with a gas-to-dust mass ratio of 100 \citep{bec90}. 
$\beta=2$ was fixed to keep consistent with that used in studies of \her\ observations toward other star-forming filaments \egcite{pal13,cox16}.
To reveal small-scale dense structures (i.e., clumps/cores), we made a high-resolution $N_{\rm H_2}$ map using the $35\arcsec$-resolution 
$T_{\rm dust}$ map to convert the 250\,\um\ intensity map to a column density map. 
This simple method has already been demonstrated to be feasible in creating a relatively high-resolution $N_{\rm H_2}$ map \citep{stu15,stu16}. 
Figure\,\ref{fig:filMorph} presents the $35\arcsec$-resolution $T_{\rm dust}$ map, and 
the final $18\arcsec$-resolution $N_{\rm H_2}$ map.

\subsection{Filament properties}

\begin{figure*}
\centering
\includegraphics[width=3.4 in]{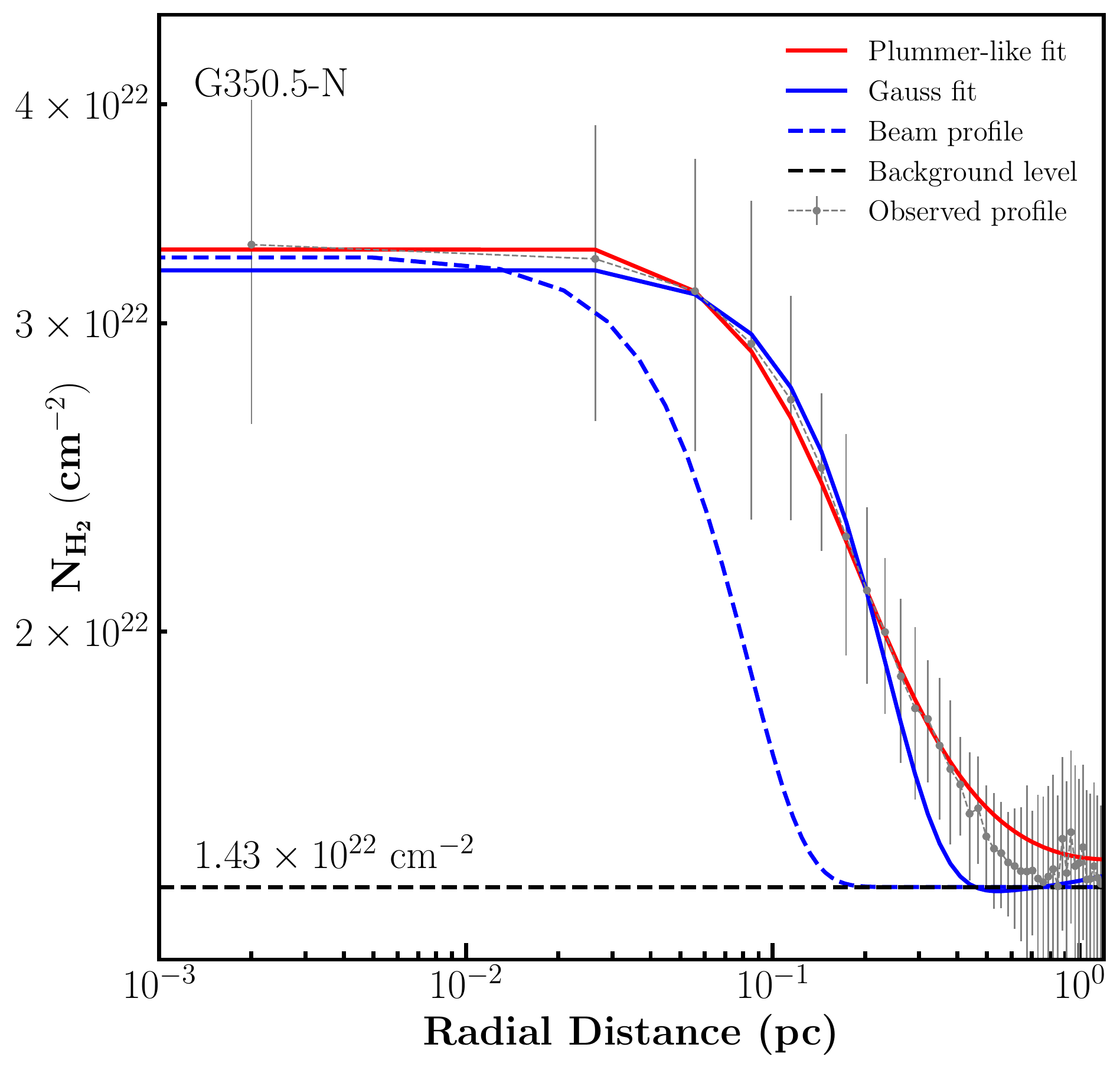}
%\hskip 0.000000005cm
\includegraphics[width=3.4 in]{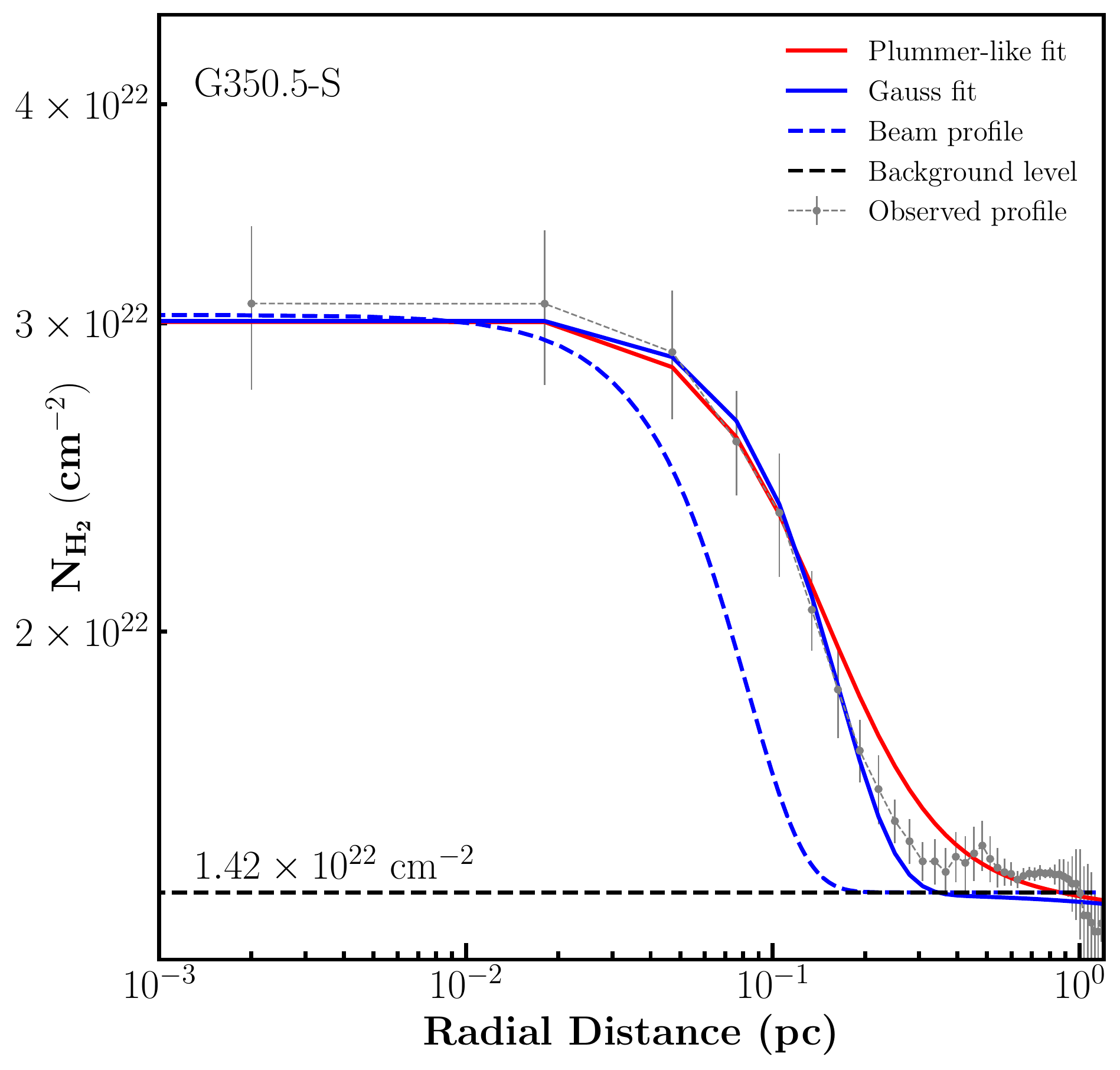}
\caption{Radial column density profiles of filaments\,G350.5-N ({\it Left}) and  G350.5-S ({\it Right}). The gray dots are 
the mean observed data where the background is 
$N_{\rm H_{2}}^{\rm bg} = 1.43 \times 10^{22}$\,\cmcm\ for \filAname, and $N_{\rm H_{2}}^{\rm bg} = 1.42 \times 10^{22}$\,\cmcm\ for \filBname. 
The red, and blue solid lines represent Plummer-like, and Gaussian fits, respectively. 
The blue dashed line corresponds to the $18\farcs2$ beam resolution.
For the filament G350.5-N, the key parameters are $R_{\rm flat}=0.15\pm0.01$\,pc, and $p=3.0\pm 0.1$ for the plummer-like fitting, 
and  ${\rm FWHM}=0.34\pm0.01$\,pc for the Gaussian fitting. For the filament G350.5-S, the corresponding fitting parameters are
$R_{\rm flat}=0.12\pm0.01$\,pc, $p=3.5\pm 0.5$, and ${\rm FWHM}=0.24\pm0.01$\,pc.}
\label{fig:fildenprof}
\end{figure*}

As shown in Fig.\,\ref{fig:filMorph}, the filament \filname\ is composed of two discontinuous filaments in both dust temperature and column density maps. 
To better delineate them, the function {\it make\_fil\_spine} built in the algorithm {\it FilFinder}\footnote{\url{https://github.com/e-koch/FilFinder}} 
was used to extract the main skeleton of the filaments from the column density map. For simplicity, {\it FilFinder} identifies the filamentary structures using an adaptive thresholding, 
which creates a mask based on local brightness (or intensity) changes in the image. The method Medial Axis Transform is then used to reduce the signal mask to a skeleton. 
This skeleton is finally pruned down to a filamentary network.  More detailed descriptions on {\it FilFinder} can be found in \citet{koc15}.
With that algorithm, two discontinuous filaments are well identified (see Fig.\,\ref{fig:filMorph}). Given the distance to \filname, the northern 
part is estimated to be $\sim 5.9$\,pc long and the southern one is $\sim 2.3$\,pc long.
The two parts are called \filAname, and \filBname\ in the following discussions, respectively. 

\filAname, and \filBname\ are observed to be associated with each other with similar systematic velocities based on
the \thco\ and \etco~(2-1) observations (see Sect.\,\ref{sec:cores}). To make an approximated calculation, we define these two discontinuous filaments 
within a column density level of $1.8\times 10^{22}$\,\cmcm\ that surrounds the majority of their masses. After 
subtracting  an approximate background emission of $1.43\times10^{22}$\,\cmcm, which was estimated from the column density profiles of the 
two filaments (see Sect.\ref{sec:profile}), we obtain the total mass,
mean column density, and mean temperature  to be $\sim 590$\,\msun, $\sim 8.2\times10^{22}$\,\cmcm, and $\sim 18.2$\,K, 
respectively, for \filAname\, and 
$\sim 95$\,\msun, $\sim 8.2\times10^{21}$\,\cmcm, and $\sim 17.7$\,K, respectively, for \filBname.

\subsubsection{Column density profile}
\label{sec:profile}
To better and quantitatively characterize the filaments, we construct the column density profiles for \filAname, and \filBname\ using the column density map
with the same procedure as adopted in previous studies 
\egcite{arz11,pal13,cox16}. In a nutshell, we first created the individual column density profile in the local tangent direction for each pixel along the crest 
(i.e., the red curve in Fig.\,\ref{fig:filMorph})
of the filament. The mean column density profile was then derived by averaging the individual profiles for all pixels. The resulting column density profiles for 
the two filaments \filAname, and \filBname\ are shown in Fig.\,\ref{fig:fildenprof}, where the gray dots with error bars stand for the mean measured column densities and 
the errors come from the standard deviation. 

The Plummer-like function is commonly used to describe the column density profile of a filament. According to \citet{cox16}, the form of the Plummer-like function can be simplified as
below:
\begin{equation}\label{eq:luminosity}\label{eq:plummer}
N_{\rm p}(r) = \frac{N_{\rm H_{2}}^0}{{\rm cos}(i) \ \left(1+(r/R_{\rm flat})^2\right)^\frac{p-1}{2}},
\end{equation}
where $N_{\rm H_{2}}^0$ is the central column density, $R_{\rm flat}$ is the flattening radius, $p$ is the index of a power-law fall off in column density beyond $R_{\rm flat}$,
and $i$ is the inclination angle of the filament to the plane of the sky, assumed to be 0 degrees here. Before the Plummer-like function fitting, we estimated the background column density
level fitting a first order polynomial to the selected radius range (1\,pc~$<|r|<2$\,pc) where the observed column density profile appears to be in a constant level for the two filaments. This method
gives rise to $N_{\rm H_{2}}^{\rm bg} = 1.43 \times 10^{22}$\,\cmcm\ for \filAname, and $N_{\rm H_{2}}^{\rm bg} = 1.42 \times 10^{22}$\,\cmcm\ for \filBname, respectively.
As a first-order approximation, the mean value $\overline{N_{\rm H_{2}}^{\rm bg}}=1.43\times10^{22}$\,\cmcm\ is used as background emission for the whole \filname\ cloud.
 After the background subtraction to the observed mean column density profile, 
the Plummer-like function fits were made in a radius range of $|r|<0.5$\,pc using the MPFIT non-linear least-squares fitting programme \citep{mar09}, which can propagate the uncertainties in the mean 
observed column 
density profile into estimating the uncertainties in the derived parameters. The radius range $|r|<0.5$\,pc is chosen since the majority of the densities of the two filaments
in that range are above the background level.
The best fits as indicated with red lines
in Fig.\,\ref{fig:fildenprof} yield the parameters $R_{\rm flat}=0.15\pm0.01$\,pc, and $p=3.0\pm 0.1$ for \filAname, as well as $R_{\rm flat}=0.12\pm0.01$\,pc, and $p=3.5\pm 0.5$
for \filBname. 

As mentioned in Sect.\,\ref{sec:intro}, the inner width (FWHM) of the filament is important 
for constraining the density profile and filament formation scenarios. 
By fitting a single gaussian function (i.e, the blue solid line in Fig.\,\ref{fig:fildenprof}) to the observed column density profile in an inner range of $|r|<0.5$\,pc, in which
the inner part of the two filaments appears to reside as shown in Fig.\,\ref{fig:fildenprof},
the inner widths  are derived to be $0.34 \pm0.01$\,pc for \filAname, and $0.24\pm0.01$\,pc for \filBname, which correspond to 
the deconvolved FHWMs $0.32 \pm0.01$\,pc, and $0.21\pm0.01$\,pc, respectively, given a beam size of 0.12\,pc.

The fitted parameters $R_{\rm flat}$, $p$, and FHWM should be regarded as approximate for the following reasons. The radial profile in Fig.\,\ref{fig:fildenprof} 
shows a very flat and wide inner profile for both filaments, implying that the central crest may not be resolved. Fitting a Gaussian or a Plummer-like function 
to a flattened profile would then show a larger width (i.e., $R_{\rm flat}$, FWHM) than that of a more resolved filament. The Plummer-like
 parameters (e.g., $R_{\rm flat}$, $p$) are strongly dependent on each other \egcite{mal12,juv12,smi14}.
That is, if $R_{\rm flat}$ is wider than it should be, it would in turn directly affect the fitted $p$ value. It is, therefore, worthwhile to carry out in the future 
high-resolution continuum observations, which would be very helpful for investigating the 
detailed density profiles of both \filAname\ and \filBname.

\subsubsection{Line mass distribution}
\label{sec:mline}
\begin{figure*}
\centering
\includegraphics[width=3.4 in]{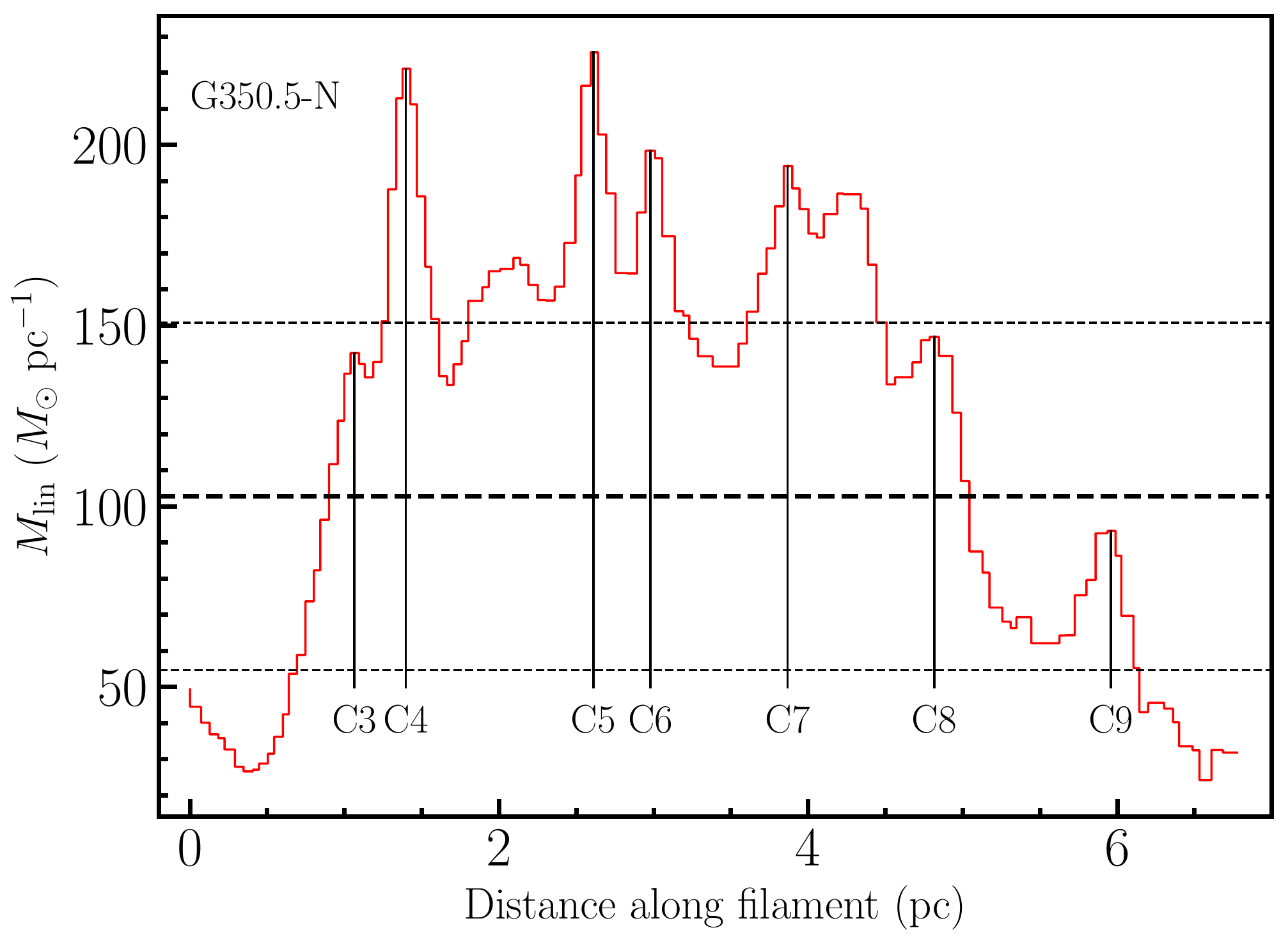}
\includegraphics[width=3.3 in]{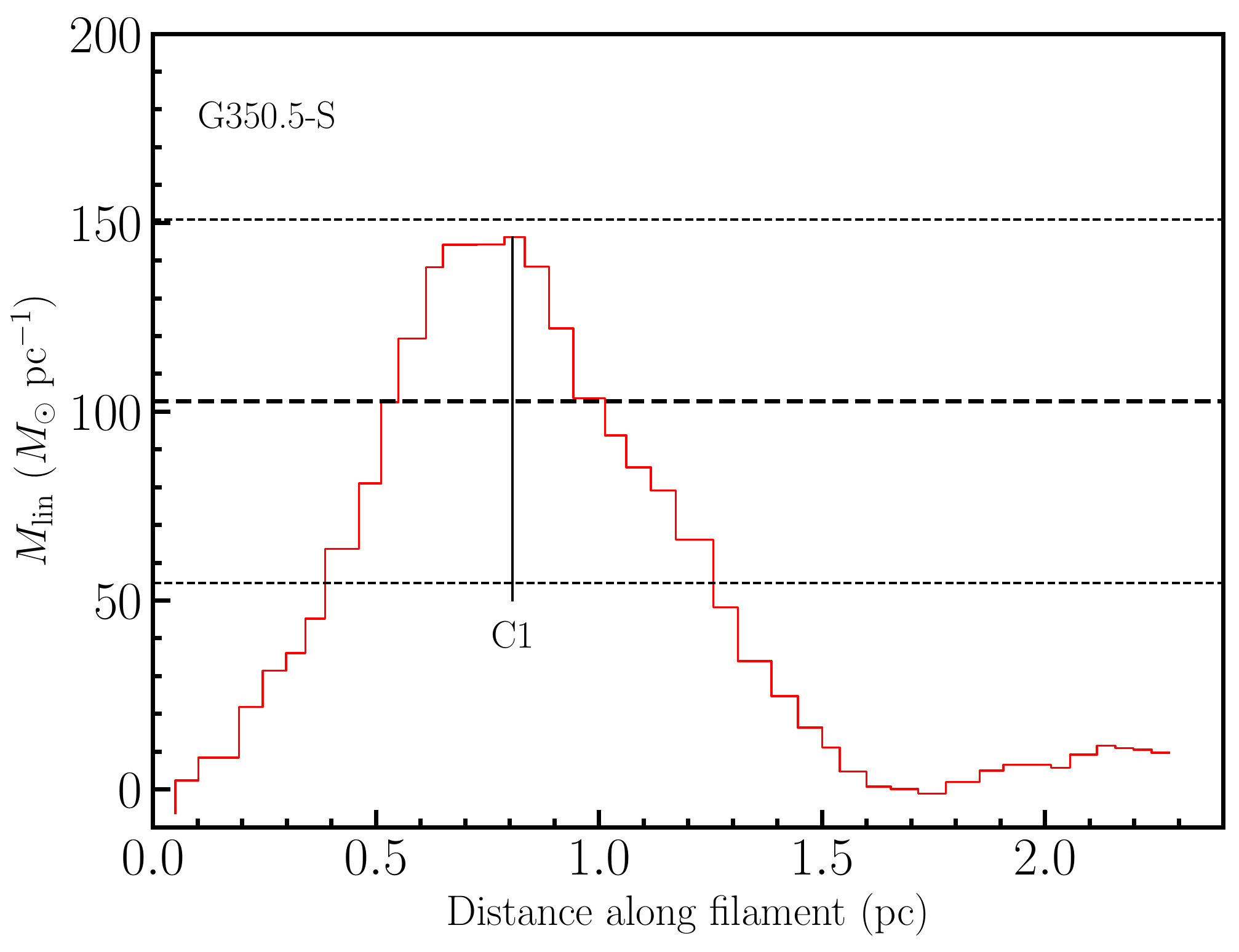}
\caption{ Mass per unit length (\mline) along the filament main axis for \filAname\ ({\it Left}), and \filBname\ ({\it Right}). The thick dashed lines show the critical 
\mline~$\sim 102$\,\lmsun\ for both \filAname\ and \filBname, and the thin dashed lines an uncertainty of $\sim 48$\,\lmsun. 
Several peaks are related to dense cores (see Fig.\,\ref{fig:filcores}) as indicated by 
vertical lines. It can bee seen that most parts of the two filaments have $M_{\rm line}>M_{\rm line}^{\rm crit}$, suggesting fragmentation
processes taking place on both filaments due to gravitational instability.
}
\label{fig:filLineMass}
\end{figure*}

One of the advantages of building the column density profile is to use it to derive the mass per unit length ($M_{\rm line}$) along the main axis of the filament. 
We calculated $M_{\rm line}$ for each position along the crest of the filament by integrating the observed column density profile over its outer width (i.e., $|r|<0.5$\,pc). 
The resulting  $M_{\rm line}$ distributions for \filAname, and \filBname\ are displayed in  Fig.\,\ref{fig:filLineMass}, where several sharp $M_{\rm line}$ peaks are related to 
dense cores (see Sect.\,\ref{sec:cores}) and the dashed horizontal lines represent the critical mass per unit length ($M_{\rm line}^{\rm crit}$). 
According to \citet{inu97}, $M_{\rm line}^{\rm crit}$ is defined as $2\sigma_{\rm th}^2/G$ for an unmagnetized isothermal filament, where $\sigma_{\rm th}$ corresponds to
the thermal sound speed and $G$ is the gravitational constant. Given the mean dust temperature of each filament (i.e., 18.2\,K for \filAname, and 17.7\,K for \filBname, 
see Sect.\,\ref{sec:dustmaps}), the critical line masses are $\sim 30$\,\lmsun\ for \filAname, and $\sim 29$\,\lmsun\ for \filBname. 
Note that we neglected the non-thermal support in the above calculation. Therefore, using the actual velocity dispersion, $0.47\pm 0.11$\,\vel\ (see Sect.\,\ref{sec:dis-densityproile}) 
instead of the thermal sound speed, we obtain the critical value $M_{\rm line}^{\rm crit} \sim 102\pm48$\,\lmsun\ for both \filAname\ and \filBname.
As shown in Fig.\,\ref{fig:filLineMass}, most parts of the two filaments have $M_{\rm line}>M_{\rm line}^{\rm crit}$. This suggests that both filaments  
could be supercritical on the process of fragmentation into prestellar dense cores owing to gravitational instability, which is well consistent with 
the picture of dense cores detected in the two filaments (see Sect.\,\ref{sec:cores}).  In addition, we calculated the total mass for these 
two filaments by integrating the $M_{\rm line}$ along the main axis of each filament, resulting in $\sim 810$\,\msun\ for \filAname, 
and $\sim 110$\,\msun\ for \filBname, respectively.
These values are in agreement with those estimated before from the column densities as described in Sect.\,\ref{sec:dustmaps}. 

\subsection{Dense cores}
\label{sec:cores}

\begin{figure*}
\centering
\includegraphics[width=3.4 in, height=3.0 in]{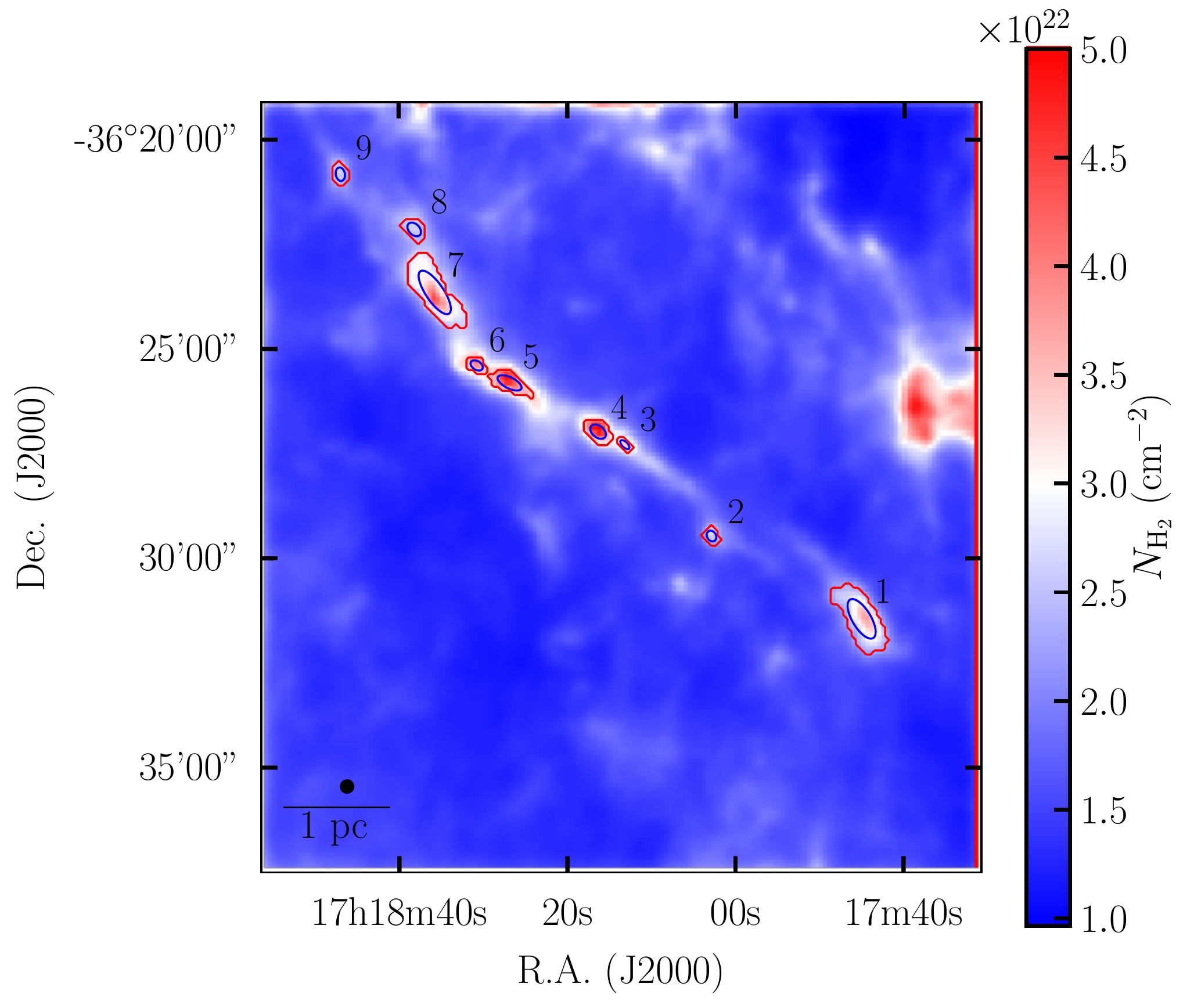}
\includegraphics[width=3.4 in, height=2.8 in]{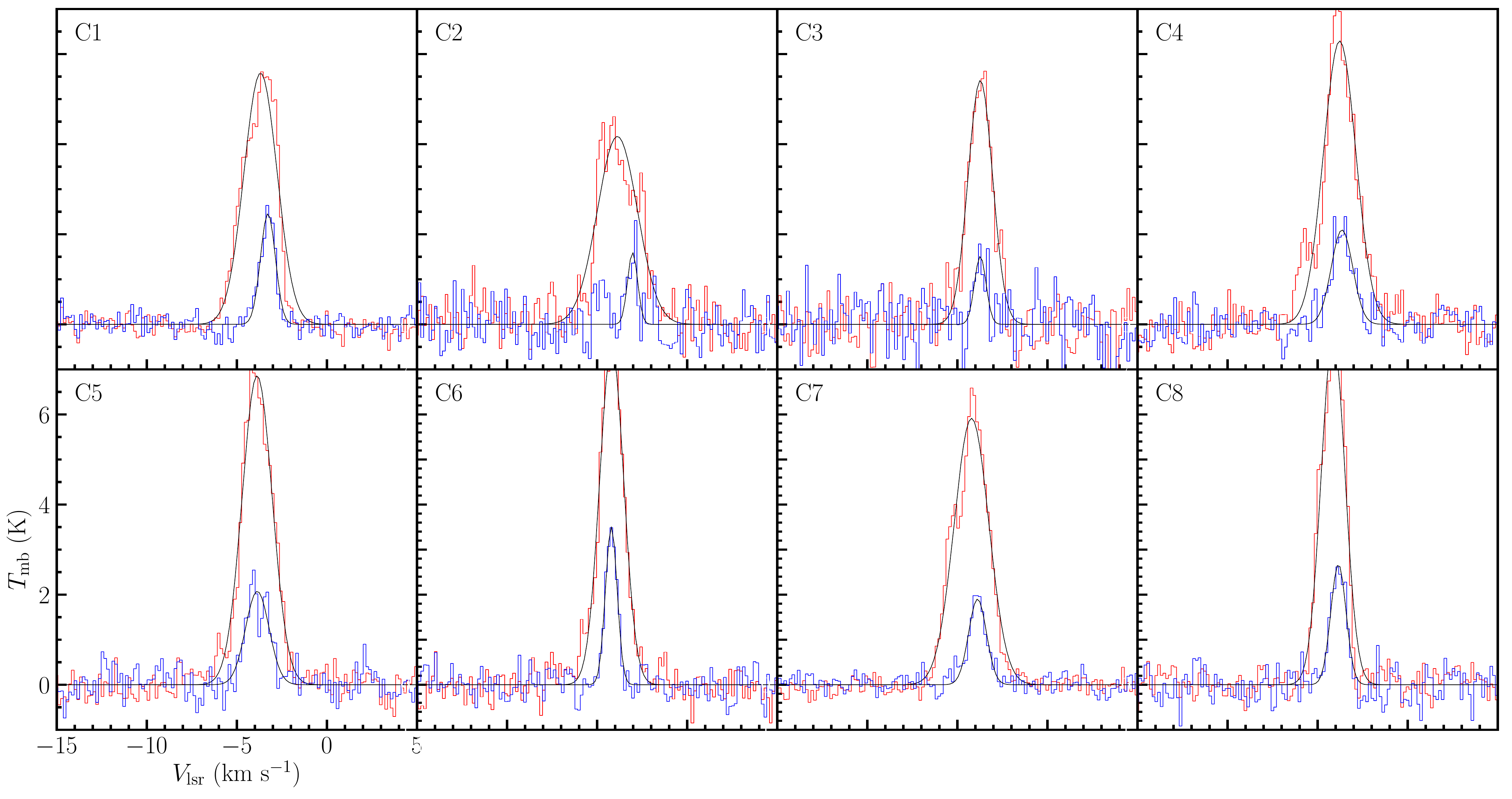}
\caption{ {\it Left:} nine identified dense cores. The column density map (color scale) is 
shown on the log scale.
{\it Right:} spectra of $J=2-1$ of \thco\ (in red) and \etco~(in blue) for the nine dense cores. The Gaussian fitting is indicated with the black curve.}
\label{fig:filcores}
\end{figure*}

Nine dense cores were identified on the filaments using the column density map (see Fig.\,\ref{fig:filMorph}).  
We made use of the Python package 
{\it astrodendro}\footnote{\url{http://dendrograms.org/}} to identify potential dense entities within the main structure of the filaments. 
As requested by the source extraction algorithm, 
a column density of $2.1\times10^{22}$\,\cmcm\ was input to be a start level to construct the dendrogram and a threshold of 9 pixels ($>18\arcsec$) was imposed to ensure a leaf to be resolvable
under current resolution.
As a result, nine dense cores are identified as shown in Fig.\,\ref{fig:filcores} and their corresponding
parameters are summarized in Table\,\ref{tbl:dustprop}, including the identity number, coordinates, effective beam-deconvolved radius, 
dust temperature, column density, number density, and core mass.

Figure\,\ref{fig:filcores} presents the spectra of the transition $J=2-1$ of \thco\ and \etco\ for all identified dense cores except for the C9 core not covered by
our current APEX observations. It can be seen that emission of both \thco\ and \etco~(2-1) is strong enough to be detected in all of the eight dense cores.
Fitting a single-Gaussian function to the observed spectra with the software GILDAS, we retrieved the corresponding observed parameters for each core, 
including the main-beam temperature ($T_{\rm mb}$), peak velocity ($V_{\rm lsr}$), and velocity dispersion ($\sigma$) for  both spectra (see Table\,\ref{tbl:lineprop}).
The peak velocities derived from both \thco\ and \etco~(2-1) are consistent with each other within 0.4\,\vel\ in all eight cores but
the core C2, suggesting that both tracers basically trace the same region. The C2 core appears to have two velocity components in \etco~(2-1) emission, 
one of which is detectable and the other one is rather noisy (i.e., 0.42\,K, three times less than the rms level).
Moreover, the systematic kinetic velocity of the two filaments is estimated to be $-3.9$\,\vel, a median value of the peak velocities of the eight cores, which has been adopted
in calculating the kinematic distance of the cloud \filname\ (see Sect.\,\ref{sec:intro}).

\begin{table*}
\centering
\caption{Dust properties of dense cores from \her\ observations}
\label{tbl:dustprop}
\resizebox{16cm}{!}{
\begin{tabular}{crrrrrrrr}
\hline\hline
ID & $R.A.$ & $Dec.$ & $R_{\mathrm{dec}}^{\rm eff,a}$ & $T_\mathrm{d}^b$ & $N_{\rm H_2}^c$ & $n_{_{\rm H_2}}^d$ & $M_\mathrm{core}^e$  & $M_\mathrm{BE}$  \\
   & J2000  & J2000  & pc                           & K              & $10^{22}$\,cm$^{-2}$ & $10^4$\,cm$^{-3}$  & \msun\             & \msun\           \\
\hline
   1  &  17:17:45.36  &  -36:31:48.0  &    0.24  $\pm$    0.02  &    17.0  $\pm$     0.3  &     1.1  $\pm$     0.4  &     1.4  $\pm$     0.1  &    56.0  $\pm$    10.6  &    23.4  $\pm$     2.3  \\ 
   2  &  17:18:03.36  &  -36:29:24.0  &    0.05  $\pm$    0.01  &    18.2  $\pm$     0.1  &     1.0  $\pm$     0.2  &    12.3  $\pm$     1.2  &     4.4  $\pm$     0.8  &     2.3  $\pm$     0.2  \\ 
   3  &  17:18:13.68  &  -36:27:36.0  &    0.03  $\pm$    0.01  &    17.7  $\pm$     0.1  &     1.8  $\pm$     0.1  &    69.8  $\pm$     6.6  &     5.9  $\pm$     1.1  &     1.8  $\pm$     0.2  \\ 
   4  &  17:18:16.80  &  -36:27:00.0  &    0.11  $\pm$    0.01  &    17.4  $\pm$     0.1  &     2.1  $\pm$     0.6  &     6.9  $\pm$     0.6  &    27.0  $\pm$     5.1  &    23.0  $\pm$     2.3  \\ 
   5  &  17:18:27.36  &  -36:25:48.0  &    0.13  $\pm$    0.01  &    17.2  $\pm$     0.3  &     2.1  $\pm$     0.5  &     5.7  $\pm$     0.5  &    34.5  $\pm$     6.5  &    31.9  $\pm$     3.2  \\ 
   6  &  17:18:31.20  &  -36:25:12.0  &    0.07  $\pm$    0.01  &    17.2  $\pm$     0.1  &     2.2  $\pm$     0.2  &    14.7  $\pm$     1.4  &    13.0  $\pm$     2.5  &     4.2  $\pm$     0.4  \\ 
   7  &  17:18:36.24  &  -36:24:00.0  &    0.25  $\pm$    0.02  &    17.7  $\pm$     0.3  &     1.4  $\pm$     0.5  &     1.7  $\pm$     0.2  &    73.7  $\pm$    13.9  &    36.2  $\pm$     3.6  \\ 
   8  &  17:18:38.64  &  -36:22:12.0  &    0.09  $\pm$    0.01  &    18.2  $\pm$     0.2  &     1.0  $\pm$     0.1  &     4.2  $\pm$     0.4  &     8.9  $\pm$     1.7  &     9.2  $\pm$     0.9  \\ 
   9  &  17:18:47.52  &  -36:21:00.0  &    0.07  $\pm$    0.01  &    17.9  $\pm$     0.1  &     0.9  $\pm$     0.1  &     5.8  $\pm$     0.5  &     6.0  $\pm$     1.1  &     0.0  $\pm$     0.0  \\ 

\hline
\end{tabular}

}

\hbox{$^{\rm a}$ The beam deconvolved radius.}
\hbox{$^{\rm b}$ The average dust temperature over the size of the core.}
\hbox{$^{\rm c}$ The average dust column density over the size of the core.}
\hbox{$^{\rm d}$ $n_{\rm H_{2}} = N_{\rm H_{2}}/2 R_{\rm dec}^{\rm eff}$.}
\hbox{$^{\rm e}$ $M_\mathrm{core} = \frac{4\pi}{3} R_{\rm{dec}}^{3} n_{_{\rm H_{2}}} \mu_{_{\rm H_{2}}} m_{\rm H}$, where $\mu_{_{\rm H_{2}}} = 2.8$ is 
the mean atomic weight of gas.}

\end{table*}

\begin{table}
\centering
\caption{Properties of dense cores from line observations}
\label{tbl:lineprop}
\resizebox{9cm}{!}{
\begin{tabular}{crrrrrrrr}
\hline\hline
   & \multicolumn{3}{c}{$^{13}{\rm CO}$} & \multicolumn{3}{c}{${\rm C^{18}O}$} \\
\cline{2-4} \cline{5-7} 
ID & $T_{\rm mb}^{a}$ & $V_{\rm LSR}^{b}$ & $\sigma^{c}$ & $T_{\rm mb}$ & $V_{\rm LSR}$ & $\sigma$ \\
   & K            & \vel\         &  \vel\   & K            & \vel\         & \vel\    \\
\hline
   1  &    5.6  $\pm$    0.1  &   -3.7  $\pm$    0.1  &    0.9  $\pm$    0.1  &    2.5  $\pm$    0.1  &   -3.3  $\pm$    0.2  &    0.4  $\pm$    0.1  \\ 
   2  &    4.2  $\pm$    0.1  &   -3.9  $\pm$    0.2  &    1.1  $\pm$    0.2  &    1.6  $\pm$    0.3  &   -3.0  $\pm$    0.1  &    0.3  $\pm$    0.1  \\ 
   3  &    5.4  $\pm$    0.1  &   -3.7  $\pm$    0.1  &    0.7  $\pm$    0.1  &    1.5  $\pm$    0.3  &   -3.7  $\pm$    0.1  &    0.3  $\pm$    0.1  \\ 
   4  &    6.3  $\pm$    0.1  &   -3.8  $\pm$    0.1  &    0.9  $\pm$    0.1  &    2.1  $\pm$    0.1  &   -3.6  $\pm$    0.2  &    0.6  $\pm$    0.2  \\ 
   5  &    6.9  $\pm$    0.1  &   -3.9  $\pm$    0.3  &    0.8  $\pm$    0.1  &    2.1  $\pm$    0.1  &   -3.9  $\pm$    0.1  &    0.7  $\pm$    0.1  \\ 
   6  &    7.5  $\pm$    0.2  &   -4.2  $\pm$    0.2  &    0.7  $\pm$    0.1  &    3.5  $\pm$    0.2  &   -4.2  $\pm$    0.1  &    0.3  $\pm$    0.1  \\ 
   7  &    5.9  $\pm$    0.1  &   -4.2  $\pm$    0.1  &    1.0  $\pm$    0.2  &    1.9  $\pm$    0.1  &   -3.9  $\pm$    0.3  &    0.5  $\pm$    0.1  \\ 
   8  &    7.7  $\pm$    0.1  &   -4.1  $\pm$    0.1  &    0.7  $\pm$    0.1  &    2.7  $\pm$    0.2  &   -3.8  $\pm$    0.1  &    0.4  $\pm$    0.1  \\ 

\hline
\end{tabular}

}

\hbox{$^{\rm a}$ The main beam temperature.}

\hbox{$^{\rm b}$ The systematic velocity.}

\hbox{$^{\rm c}$ The velocity dispersion.}

\end{table}

\section{Discussions}
\label{sec:conc}

\subsection{Column density profile analysis}
\label{sec:dis-densityproile}

The mean column density profiles of \filAname, and \filBname\ are described in a Plummer-like function with a power law index of $p \sim 3$ (see Fig.\,\ref{fig:fildenprof}).
This index is in a reasonable previously observed range. For example,
a range of $p \sim 1.5 - 3$ has been reported in several filaments, such as $p=1.5-2.5$ observed towards the filaments in the IC5146 cloud \citep{arz11}, $p=2.0\pm0.4$
in the Taurus B211/3 filament \citep{pal13}, $p=2.2\pm0.3$ in high-mass star-forming ridge of the NGC\,6334 cloud \citep{and16}, $p=2.7\pm0.2$ in the cloud Vela\,C \citep{hil12},
$p=2.7-3.4$ in L1517 \citep{hac11}, and $p=3$ in L1495 \citep{taf15}. Theoretically, the radial equilibrium of filamentary molecular
clouds can be described by treating filaments as isothermal cylinders and using either pure hydrostatic models \citep{ost64} or
magnetohydrodynamic models \egcite{fie00}. The former models can lead to a typical density profile of $\rho \propto r^{-4}$ and the latter ones can
give a characteristic density profile approaching to $\rho \propto r^{-2}$. From the observation point of view, the theoretical models with magnetic fields
can produce more realistic density profiles than those without magnetic fields based on at least two facts: 1) most of observed column density profiles are shallower than predicted for a pure 
gravitationally bound filament without magnetic fields \citep{ost64}, and 2) magnetic fields have already been observed to be important over largely dynamical
scales from 100\,pc to 0.01\,pc \egcite{lih13,lih14,lih15}. Filaments \filAname\ and \filBname\ have steeper column density profiles than those expected
for the case of a magnetized isothermal cylinder filament (i.e., $\rho \propto r^{-2}$). Keeping in mind the uncertainties of the fitted $p$ (see Sect.\,\ref{sec:profile}),
we suggest that the steeper column density profiles of both \filAname, and \filBname\ might be a result of the
radially gravitational contraction of the filament, which is consistent with the supercritical mass per unit length of both filaments, an indicator of filament 
fragmentation induced by gravitational instability (see Fig.\,\ref{fig:filLineMass}).

\begin{figure}
\centering
\includegraphics[width=3.4 in]{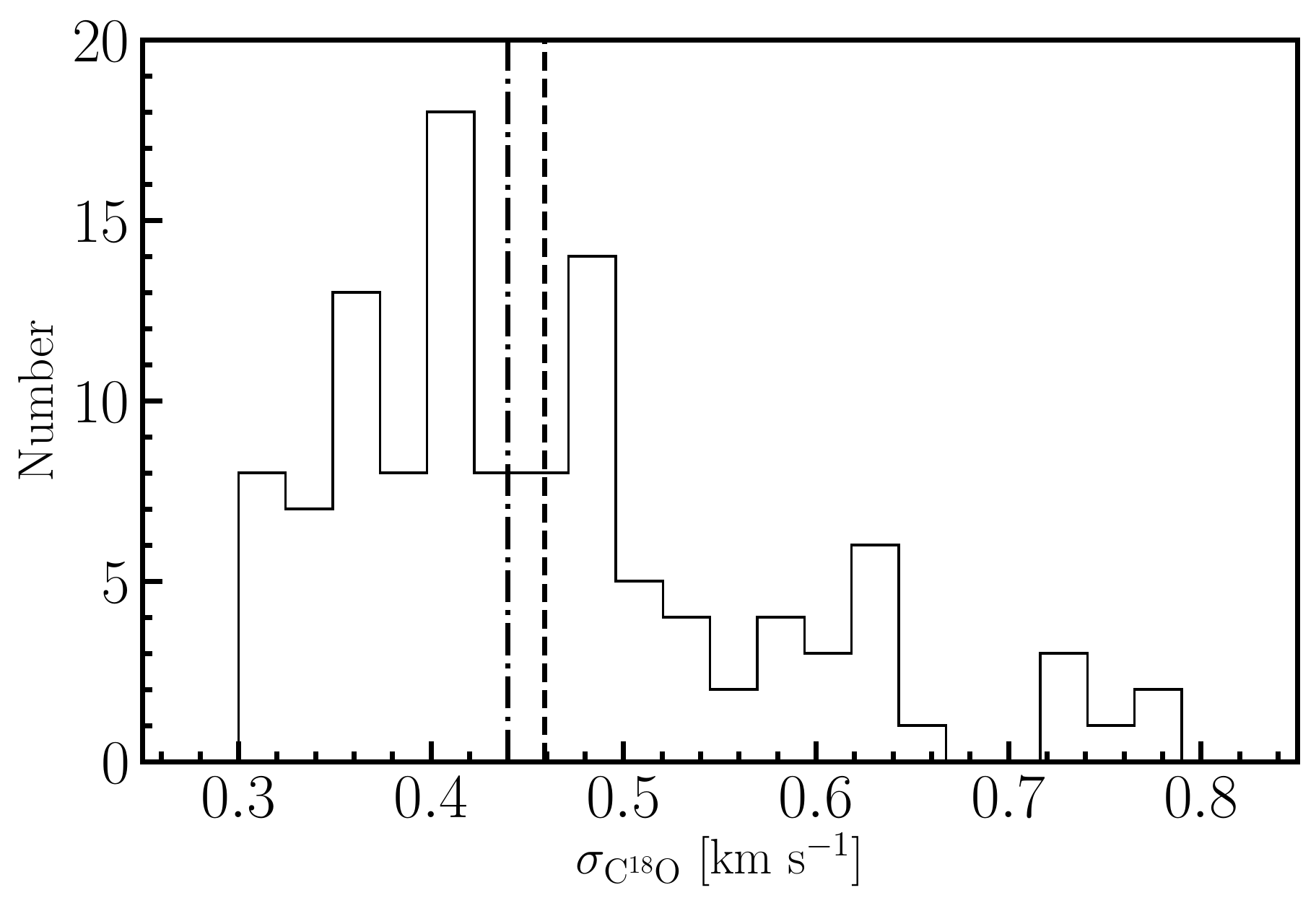}
\caption{Velocity dispersion for the positions along the main axis of the filaments \filAname, and \filBname. The velocity dispersion was measured 
by fitting a Gaussian function to the \etco~(2-1) spectrum for the positions with S/N~$>3$.  The 
dot-dashed line represents a median velocity dispersion of 0.44\,\vel\ and the dashed line indicates a mean one of $0.47$\,\vel.}
\label{fig:sigmac18o}
\end{figure}

\begin{figure}
\centering
\includegraphics[width=3.4 in]{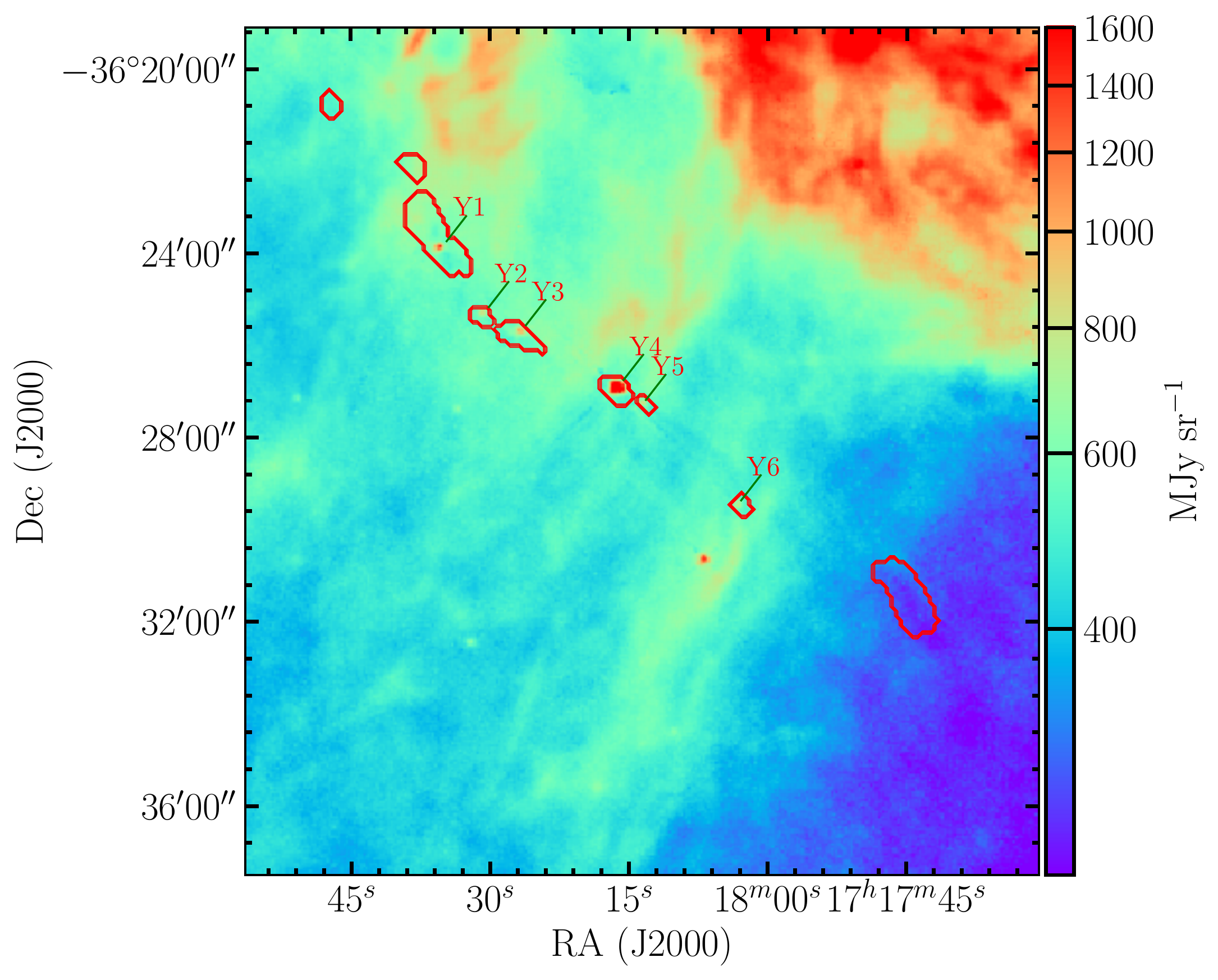}
\caption{Dust emission at 70\,\um\ overplotted with the nine dense cores identified from the high-resolution column density map.
Several bright point-like sources (i.e., Y\,1-6) associated with some of dense cores are labeled.
The main structure of G350.5 appears to be surrounded by thin perpendicular striations.
}
\label{fig:im70}
\end{figure}

Nearby filamentary clouds in the Gould Belt have been observed to be characterized with an almost constant inner width of 0.1\,pc \egcite{arz11}.
As analyzed in Sect.\,\ref{sec:profile}, the FWHMs of \filAname, and \filBname\ are $0.28 \pm0.01$\,pc, and $0.22\pm0.01$\,pc, respectively. 
These values agree with those found between 0.26 to 0.34\,pc for the filaments observed in Cygnus\,X,  and others found in the literature (see Sect.\,\ref{sec:intro}), 
suggesting that the inner width 0.1\,pc is not necessarily universal for the filaments beyond the Gould Belt. 
According to Larson's velocity dispersion-size relationship,
$\sigma_{\rm 1D}$(\vel)~$= 0.63 \times L {\rm (pc)}^{0.38}$ \citep{lar81}, the FHWMs, 0.34\,pc (0.24\,pc) for \filAname(S), lead to 
a predicted velocity dispersion of 0.42\,\vel~(0.37\,\vel). As mentioned in Sect.\,\ref{sec:profile}, the fitted FHWMs of the two filaments
are probably overestimated due to the flat density profiles caused by our current poor spatial resolution. In turn, the predicted 
velocity dispersion values could be overestimated.
Figure\,\ref{fig:sigmac18o} presents the velocity dispersion distribution of the two filaments as a whole, 
measured from the \etco~(2-1) data at the positions along the main axis. It can be seen
 that the observed mean velocity dispersion, $0.47\pm0.12$\,\vel, for the two filaments as a whole,  is slightly greater than those predicted 
 by the Larson's relationship.
This indicates that external pressures, such as the accreting flows, may produce additional turbulence into the main structure of the filaments. 
As shown in Fig.\,\ref{fig:im70}, the 70\,\um\ dust striations perpendicular to the main axis of the filaments, especially
for \filAname, are somehow in favor of the gas accretion scenario, which will be discussed further in Sect.\,\ref{sec:comparison}.
These striations can also be seen at other \her\ wavelengths from 160\,\um\ to 500\,\um\ (see Fig.\,\ref{fig:imHer}).

\subsection{Fragmentation of filaments}
\label{sec:fragment}

As discussed in Sect.\,\ref{sec:mline}, the filamentary cloud \filname\ is gravitationally unstable. Therefore,
nine dense cores detected on the main structure of the filament should
be a result of the fragmentation of \filname. The projected separations between these cores are measured to be in the range  
$\sim 0.4 - 1.7$\,pc at the cloud distance. The mean separation of these cores is $\sim 0.9\pm0.4$\,pc. The shortest two separations are 
found in two pairs of cores,  C3 and C4, C5 and C6. Regardless of these two pairs, other separations appear to be
periodic.

We take advantage of ``sausage" instability theory to examine the fragmentation of the filamentary cloud \filname\ in more details. 
The keystone of this theory is that the fragmentation of a self-gravitating 
fluid cylinder can give rise to distinct dense cores with almost periodic separations \egcite{cha53,Nag87,jac10}. This separation corresponds to
the characteristic wavelength ($\lambda_{\rm crit}$) where the instability grows the fastest.  For an incompressible fluid,
 perturbation analysis shows that the critical wavelength appears at $\lambda_{\rm crit} =  11 R_{\rm cyl}$, where $R_{\rm cyl}$
 is the cylinder's radius \egcite{cha53}. In an infinite isothermal gas cylinder, the characteristic wavelength becomes 
 $\lambda_{\rm crit} = 22 H$, where $H = c_{\rm s} (4\pi G \rho_{\rm c})^{-1/2}$ is the isothermal scale height with 
 $\rho_{\rm c}$ the central mass density along the axis of the cylinder \egcite{Nag87,inu92,jac10}.
 For a finite isothermal gas cylinder surrounded by uniform medium, $\lambda_{\rm crit}$
 is determined by the ratio of the cylinder radius to isothermal scale height, $R_{\rm cyl}/H$. If $R_{\rm cyl}/H \gg 1$, 
 the characteristic separation will approach that for the infinite isothermal gas cylinder model, $\lambda_{\rm crit} = 22 H$, but
 if $R_{\rm cyl}/H \ll 1$, it will decrease to  $\lambda_{\rm crit} = 11 H$ for the incompressible case (see \citet{jac10} for a review).

 To compare our observations with the ``sausage" instability theory, the filament cloud \filname\ is assumed to be a uniform isothermal
 cylinder. The outer radius of \filname\ is $\sim 0.5$\,pc as measured in Sect.\,\ref{sec:profile}. In addition, 
 the Plummer-like function fitting to the column density profiles
 results in a central gas volume  density of $\rho_{\rm c} = 2.7\times10^4$\,\pcmcu\ given $\rho_{\rm c} = n_{\rm c}\mu m_{\rm H}$, where 
 $n_{\rm c} = 2N_{\rm c}/\pi R_{\rm flat}$ \citep[see Eqs.\,\ref{eq:plummer}, and Eq.\,2 of][]{smi14}. 
 This central gas density represents an underestimate because  at the resolution (0.12\,pc) of our observations the central filament
  density profile may be unresolved.
 Given an average dust temperature of $\sim 18$\,K, which is derived from the temperature map (see Fig.\,\ref{fig:filMorph}),
 a thermal sound speed of 0.25\,\vel can be estimated, leading to the scale height $H \sim 0.03$\,pc.
  With only thermal turbulence considered in the above calculation, that scale height should be underestimated since the observed 
 velocity dispersion is greater than the thermal sound speed (see Fig.\,\ref{fig:sigmac18o}), which 
 suggests non-thermal turbulent pressure dominant over thermal pressure in the entire cloud \filname. With
 the typical turbulent velocity 0.47\,\vel\ for the entire cloud, one can get the 
 scale height $H \sim 0.05$\,pc, which is much less than the typical radius of \filname\ and hence follows $\lambda_{\rm crit} = 22 H \sim 1.1$\,pc.
 This theoretically predicted value is comparable to the observed separation, $\sim0.9$\,pc. Note that the underestimated central gas volume density 
 can increase the estimate of the scale height. Apart from central gas volume density, the idealized assumption of a uniform, isothermal cylinder should
 cause uncertainties on the calculation of the critical wavelength, which could explain the observed separation range $\sim 0.4 - 1.7$\,pc.

 The ``sausage" instability theory also predicts the critical maximum line mass, above which self-gravitating cylinders would collapse radially into a line. 
 That maximum line mass can be expressed as $M_{\rm line}^{\rm max} = 2 v^2/G$. As analyzed in Sect.\,\ref{sec:mline}, $M_{\rm line}^{\rm max} = \sim 102$\,\lmsun\
 is consistent with the average observed line mass $\sim 101$\,\lmsun. Therefore,  according to
 the comparisons between the observed separation, line mass and the theoretically predicted counterparts,  we suggest that the whole filamentary cloud \filname\  
 could have undergone radial collapse and fragmentation into distinct small-scale dense cores through the ``sausage" instability.

\subsection{Dense cores in fragments}

\begin{figure}
\centering
\includegraphics[width=3.4 in]{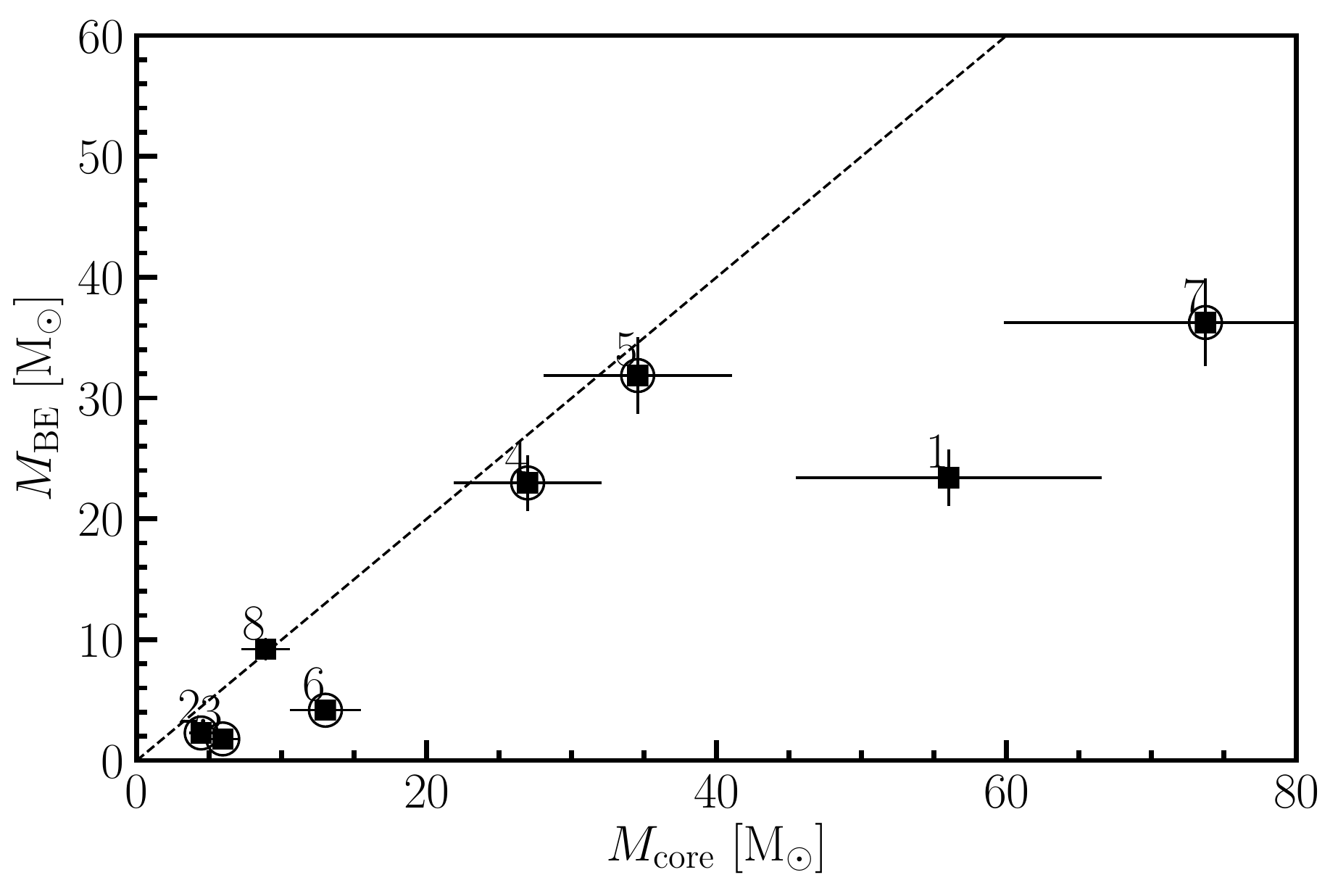}
\caption{ Comparison between core masses and the Bonnor-Ebert critical mass (derived from the core size and \etco\ velocity dispersion).  
According to the Bonnor-Ebert criterium, none of the cores in G350.5 are gravitationally stable and most have the potential to collapse. Those 
cores with YSOs are indicated with open circles.}
\label{fig:mBE}
\end{figure}

Given the filamentary cloud \filname\ fragmenting into dense cores, it is worthwhile  to assess the gravitational stability of dense cores.
We compare thermal masses with Bonnor-Ebert critical masses 
($M_{\rm BE}$) for the eight cores towards which \etco~(2-1) emission is detected using the relation from \citet{kau13}:
\begin{equation}\label{eq:virial}
M_{\rm BE} = 2.43 \frac{\sigma_{\rm v}^2R_{\rm eff}}{G}
%\alpha = \frac{M_{\rm vir}}{M_{\rm core}} = \frac{5\sigma_v^2R_{\rm eff}}{GM_{\rm core}}
\end{equation}
where $R_{\rm eff}$ is the effective radius of the dense cores, $\sigma_v$ is the velocity dispersion of gas,
and $G$ is the gravitational constant. 
The significance of Eq.\,\ref{eq:virial} is that supercritical cores with $M_{\rm BE} < M_{\rm core}$ will collapse, while subcritical cores with $M_{\rm BE} \gg M_{\rm core}$ will
expand or must be confined by additional forces \citep[e.g., external pressures][]{kau13}.

Figure\,\ref{fig:mBE} shows the core masses versus the Bonnor-Ebert critical masses for the eight cores. It can be seen that 
five out of eight cores (i.e., C1, C2, C3, C6, and C7) are gravitationally bound and the other three (i.e., C4, C5, and C8) are 
marginally bound by the gravity. The ability of those cores to gravitationally collapse is actually consistent with the picture of
star formation already taking place in some of them (see Sect.\,\ref{sec:sf}). As shown in Fig.\ref{fig:filLineMass}, 
these dense cores have conspicuous $M_{\rm line}$ corresponding to the peaks of the $M_{\rm line}$ distribution of both \filAname, and \filBname, which 
are far above the critical values around 30\,\msun.
Therefore, the eight dense cores are actually located within the supercritical filaments which are in a position to fragment and collapse to form 
new stars.

\subsection{Star formation in the filaments}
\label{sec:sf}
\begin{figure}
\centering
\includegraphics[width=3.4 in]{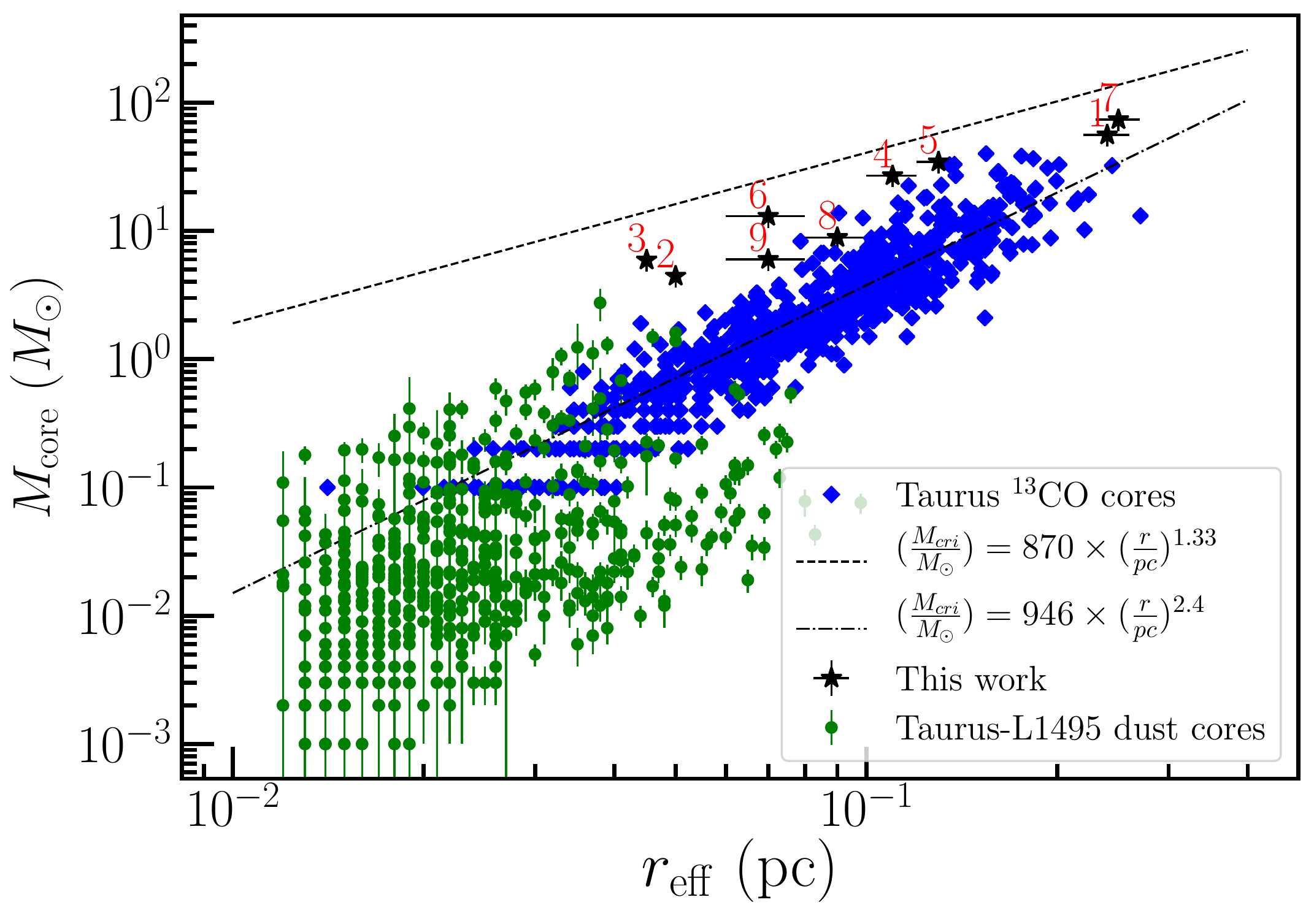}
\caption{Mass-size plot for the nine dense cores (black stars with error bars). 
The dashed line represents an empirical relation of $M(r) = 870$ \msun\ $(r/\mathrm{pc})^{1.33}$, 
which is expected to be as a threshold of forming high-mass stars \citep{kau10}.
For comparison, we add to the plot the dense dust continuum cores (green circles) in the Taurus-L1495 cloud \citep{and10,mar16}, and the 
\thco~(1-0) molecular dense cores (blue diamonds) observed in the entire Taurus cloud \citep{qia12}. The dot-dashed line 
results from the fitting to the \thco~(1-0) molecular dense cores \citep{qia12}. It can be seen that the dense
cores in G350.5 tend to be more massive than in Taurus at a fixed radius range. }
\label{fig:sizemass}
\end{figure}

\begin{figure*}
\centering
\includegraphics[width=3.4 in]{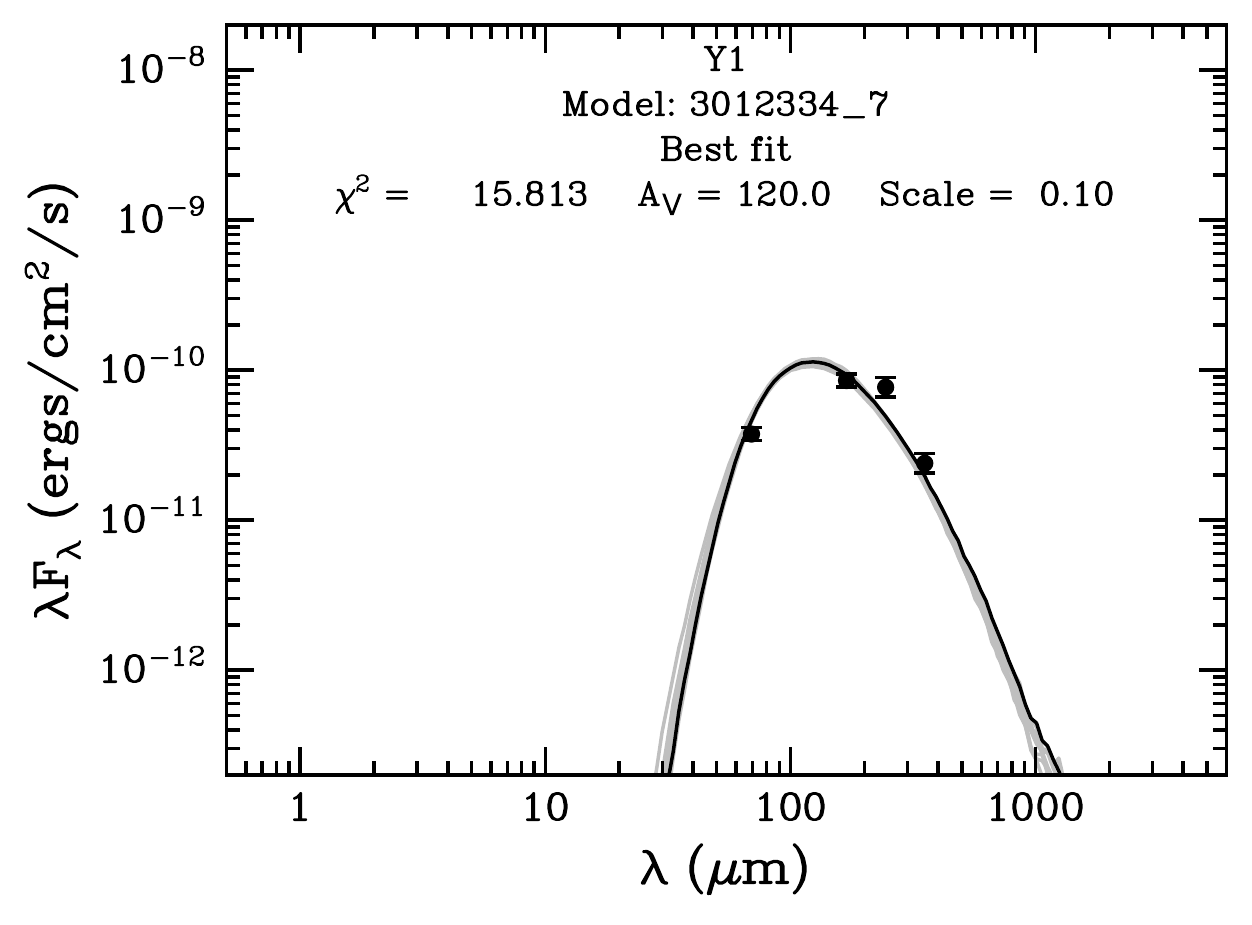}
\includegraphics[width=3.4 in]{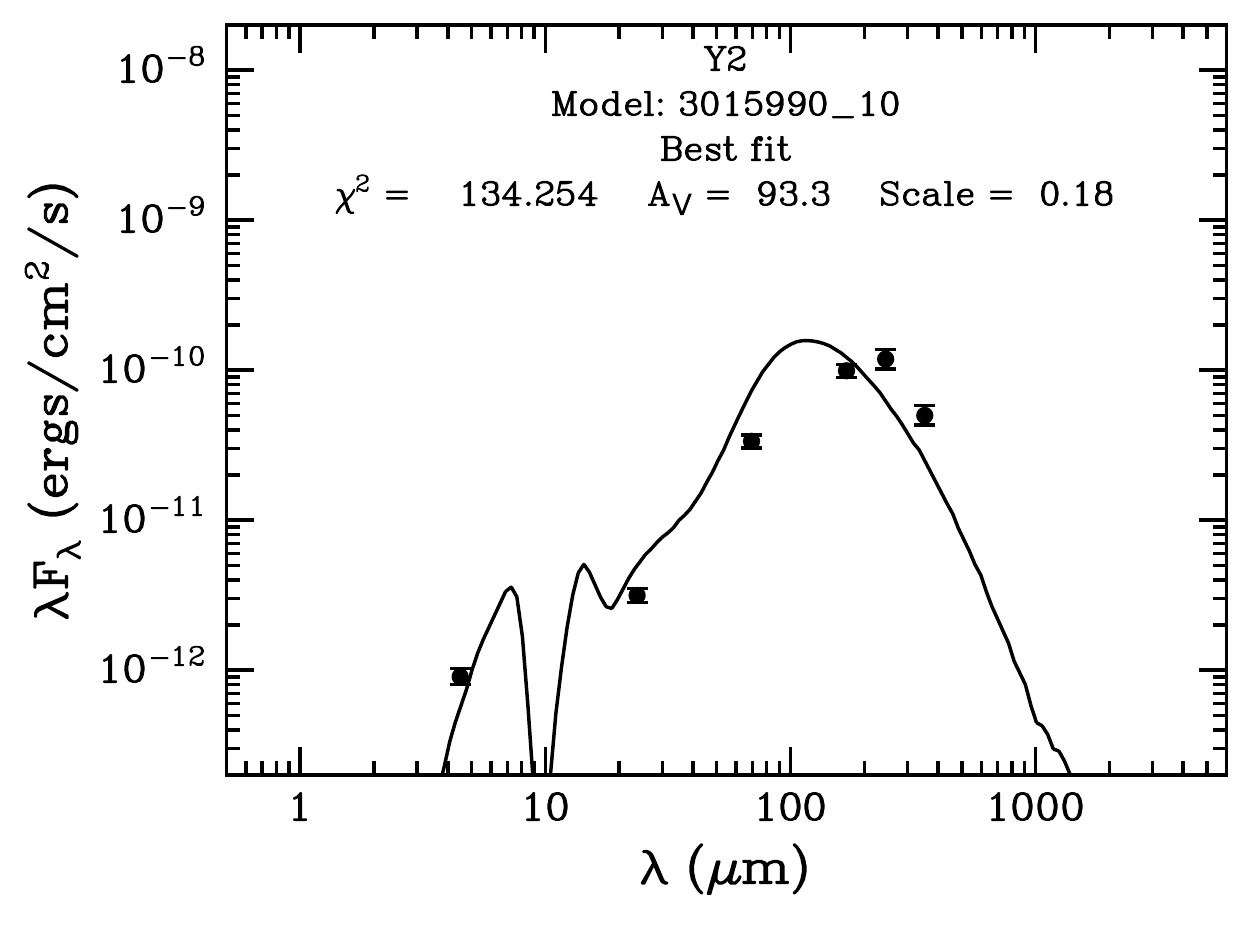}
\hskip 0.000000005cm
\includegraphics[width=3.4 in]{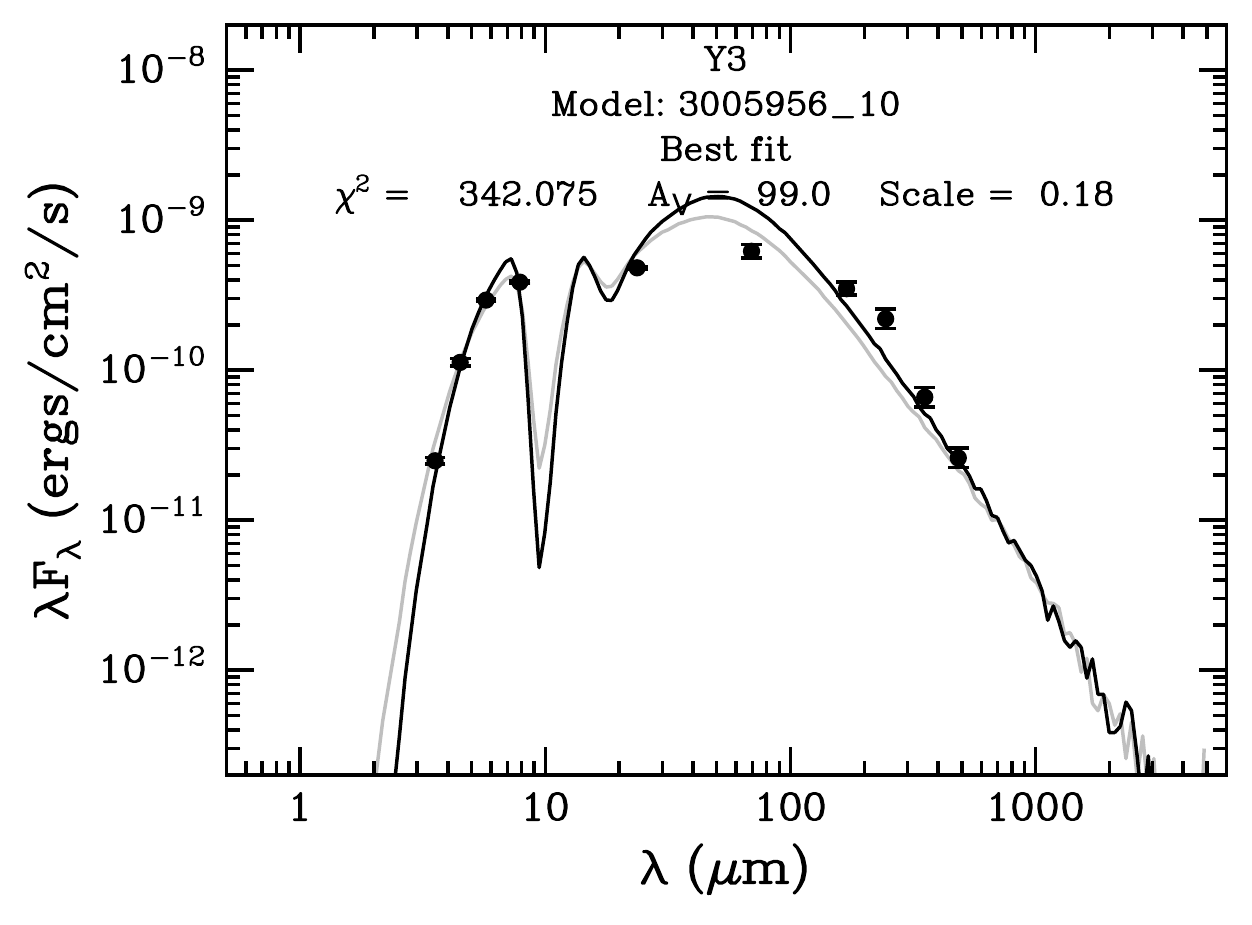}
\includegraphics[width=3.4 in]{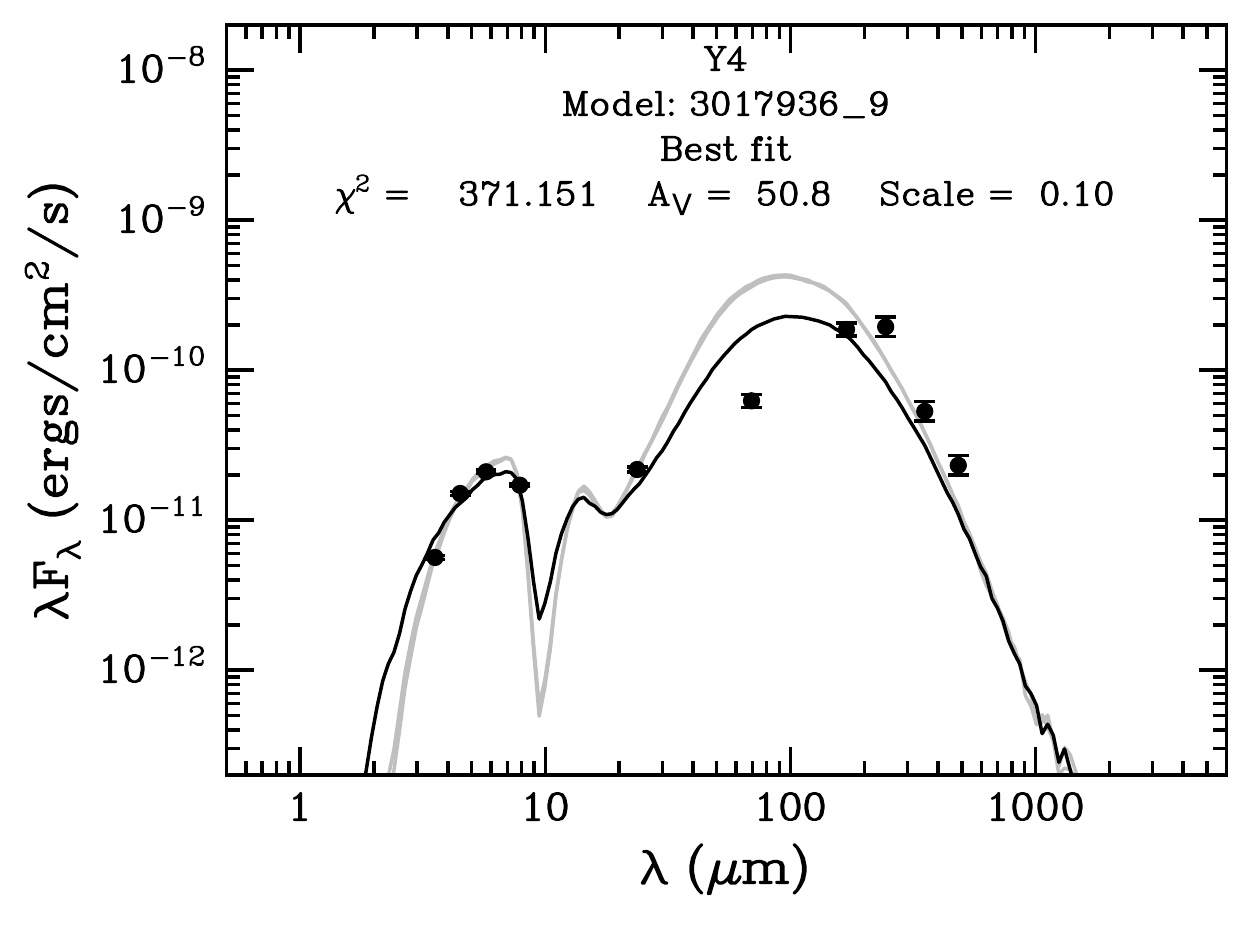}
\hskip 0.000000005cm
\includegraphics[width=3.4 in]{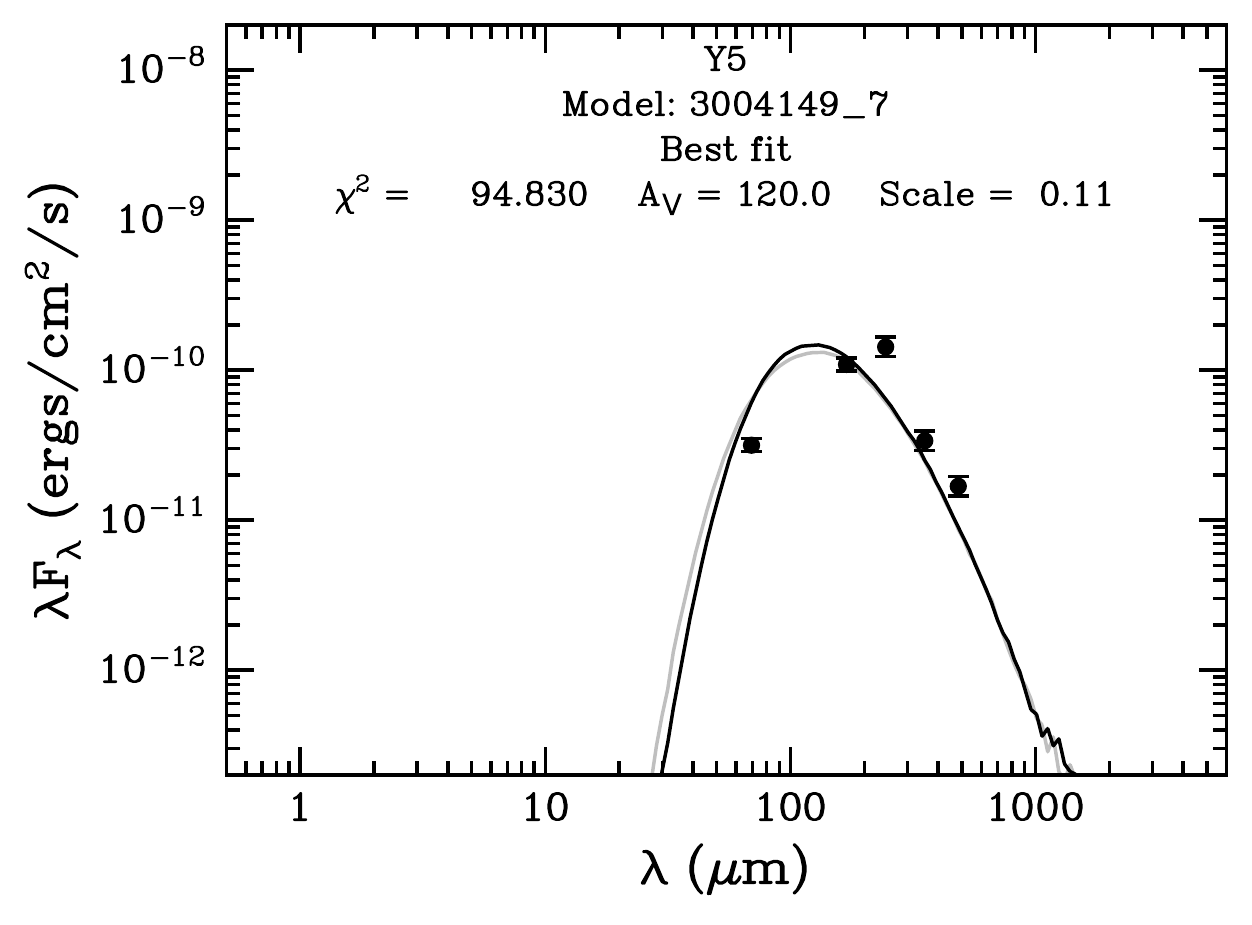}
\includegraphics[width=3.4 in]{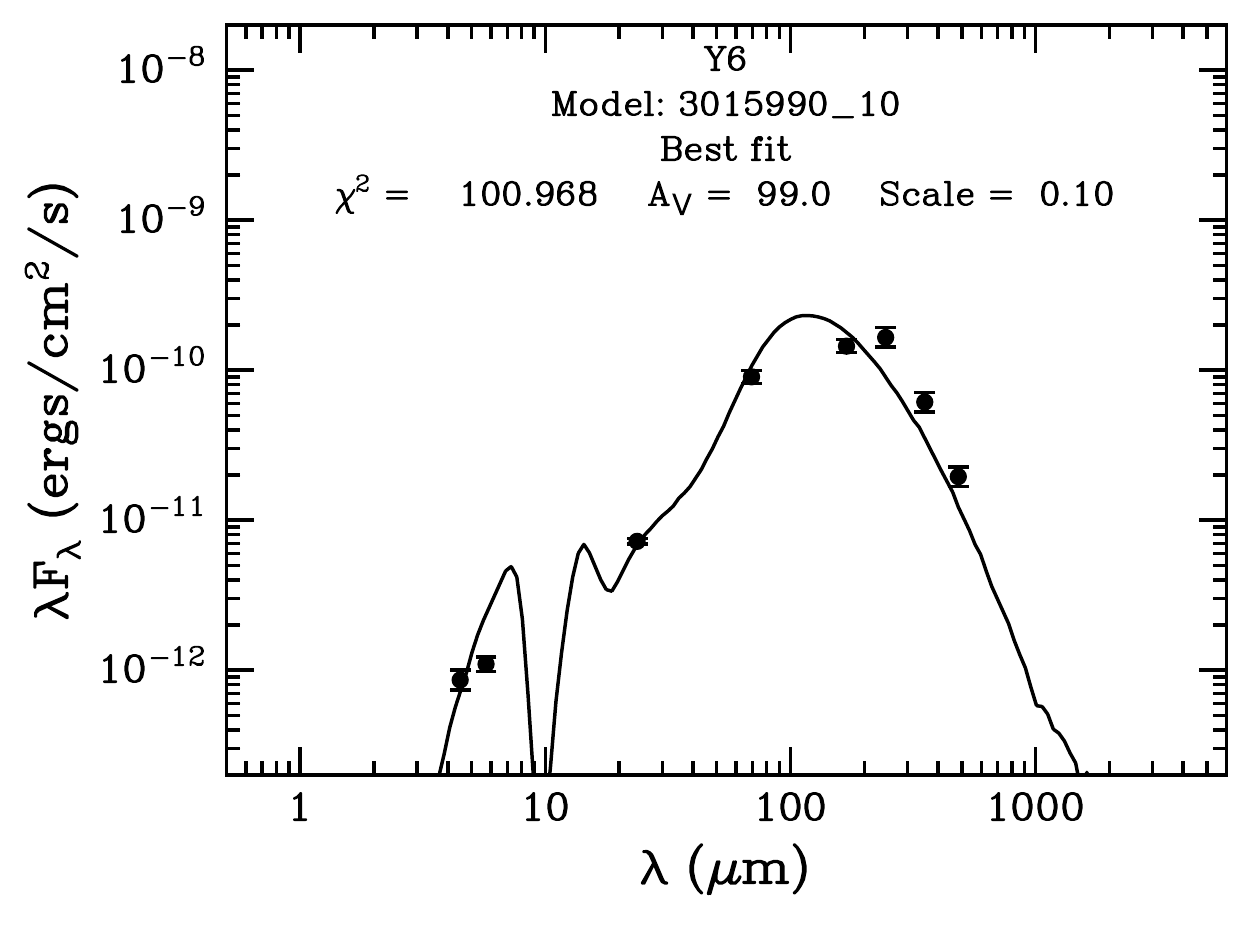}
\caption{ SED fitting plots for six candidate YSOs. The filled circles and triangle symbolize the input photometry fluxes. 
The black line shows the best fit. The gray lines show a set of fits satisfying $\chi^2-\chi_{\rm best}^2 < 3 N_{\rm data}$. 
}
\label{fig:YSOfit}
\end{figure*}

\begin{table*}
\centering
\caption{Derived parameters of young stellar objects}
\label{tbl:YSO}
\resizebox{18cm}{!}{
\begin{tabular}{crrrrrrrrrrc}
\hline\hline
ID & $R.A.$ & $Dec.$ & $N_{\rm data}$ & $\chi2/N_{\rm data}$ & $A_{_{\rm v}}^a$ & $M_{*}$  & Log($\dot{M_{\rm env}}$)       & Log$(M_{\rm disk})$ & $L_{\rm bol}$ & $Stage$ & Association \\
   & J2000  & J2000  &                &                      & Mag              & $M_{\odot}$ & Log($M_{\odot}$ yr$^{-1}$)  & Log($M_{\odot}$)    & \lsun\        &         &     \\
\hline
 Y1  &  259.513  &  -36.491  &   4  &   5.0  $\pm$   5.0  &  120.0  $\pm$   0.0  &  0.7  $\pm$  0.1  &  -3.5  $\pm$  0.1  &  -1.7  $\pm$  0.5  &   10  $\pm$   2  &  0/I &  C7 \\ 
 Y2  &  259.556  &  -36.454  &   6  &  22.4  $\pm$  22.1  &   93.3  $\pm$   0.0  &  0.7  $\pm$  0.0  &  -3.4  $\pm$  0.0  &  -3.0  $\pm$  0.0  &   13  $\pm$   4  &  0/I &  C6 \\ 
 Y3  &  259.568  &  -36.449  &  10  &  35.0  $\pm$  34.8  &   88.9  $\pm$  16.0  &  1.5  $\pm$  0.6  &  -4.7  $\pm$  0.1  &  -1.0  $\pm$  0.2  &  238  $\pm$  83  &  0/I &  C5 \\ 
 Y4  &  259.611  &  -36.429  &  10  &  38.2  $\pm$  38.0  &   65.4  $\pm$  16.7  &  2.3  $\pm$  0.9  &  -3.3  $\pm$  0.1  &  -2.5  $\pm$  0.6  &   49  $\pm$  30  &  0/I &  C4 \\ 
 Y5  &  259.628  &  -36.422  &   5  &  19.6  $\pm$  19.3  &  120.0  $\pm$   0.0  &  0.9  $\pm$  0.3  &  -3.5  $\pm$  0.1  &  -1.9  $\pm$  1.0  &   14  $\pm$   5  &  0/I &  C3 \\ 
 Y6  &  259.648  &  -36.398  &   8  &  12.6  $\pm$  12.5  &   99.0  $\pm$   0.0  &  0.7  $\pm$  0.0  &  -3.4  $\pm$  0.0  &  -3.0  $\pm$  0.0  &   13  $\pm$   4  &  0/I &  C2 \\ 

\hline
\end{tabular}

}

\begin{flushleft}
{ NOTE:} Col.\,1 is the source identity, Cols.\,2-3 the coordinates, Col.\,4
the count of data points used in the SED fitting, Col.\,5 the reduced $\chi^2$, Col.\,6 the visual extinction,
Col.\,7 the stellar mass, Col.\,8 the envelope accretion rate, 
 Col.\,9 the disk mass, Col.\,10 the bolometric luminosity, Col.\,11 the evolutionary stage, 
 Col.\,12 the spatial association of the YSO candidate with the dense core.
 The parameters in Cols\,5-10 were constrained with the normalized relative probability distribution of the well-fitted models 
 of each source \citep{yua14}, $P(\chi^{2}_{i})=\frac{e^{-(\chi^{2}_{i}-\chi^{2}_{best})/3N_{data}}}{\sum e^{-(\chi^{2}_{i}-\chi^{2}_{best})/3N_{data}}}$.
 Here, the well-fitted models are assumed to be those satisfying $(\chi^{2}-\chi^{2}_{best})/N_{data}<3$.

\end{flushleft}

\end{table*}

The filaments \filAname, and \filBname\ could be on the process of star formation as suggested by their mean line masses,
$\sim 120\pm60$\,\lmsun\ for \filAname, and $\sim 45\pm50$\,\lmsun\ for \filBname, greater than the critical line masses around 30\,\lmsun.  
Actually, \filAname, and \filBname\ do not have a large difference in 
line mass from the well-known Taurus B211/3. The filament \filAname\ is more or less twice as massive 
as B211/3 ($\sim 50$\,\mline, \citealt{pal13}) and \filBname's line mass is comparable to that of B211/3. This comparison
suggests that \filAname\ and \filBname\ could be sites of low-mass star formation. To quantitatively demonstrate this conjecture, we take advantage of an empirical mass-size relationship,
$m(r) \leq 870 M_{\odot} (r/pc)^{1.33}$, which was suggested to be an approximate threshold between low- and high-mass star formation \citep{kau10}.
 Figure\,\ref{fig:sizemass} shows the mass versus size relation for the observed nine dense cores.
One can notice that all of the nine cores are below the threshold as indicated by the dashed line, indicative of low mass star formation
most likely occurring in these dense cores. For further comparison, we collected part of dense cores in the Taurus-L1495 cloud extracted from {\it Herschel} observations 
\citep[green circles,][]{and10,mar16}, and  molecular dense cores from \thco~(2-1) observations toward the entire Taurus cloud \citep[blue diamonds,][]{qia12}. 
Figure\,\ref{fig:sizemass} shows that all of the dense cores in this work
 are basically located close to the Taurus dense cores. Therefore, the mass-size distribution of the nine dense cores suggests that they most likely form low-mass stars,
 which can also be consolidated by several low-mass YSOs detected in some of the dense cores (see below).

Six bright point sources at 70\,\um\ are found to be spatially coincident with six out of the nine cores, as shown in Fig.\,\ref{fig:im70}. 
These sources may be similar in nature to the  70\,\um\ protostars (\citet{stu13}; see also \citealt{rag12}). These authors concluded 
that protostars with very red 70\,\um\ to 24\,\um\ colors are consistent with having denser envelopes and thus are younger than 
the bulk of the Class\,0 and Class\,I protostars in Orion \citep{fur16,fis17}. To examine if these bright sources 
are YSO candidates, we conducted the fitting of spectral energy distribution (SED) with the tool developed by \citet{rob06}. This tool
 invokes a grid of 200,000 two-dimensional Monte Carlo radiation transfer models, working as a linear regression
 method to fit these models to the multi-wavelength photometry
measurements of a given source.  We first retrieved the photometry
 data matched within $5\arcsec$ from the center coordinates of each bright source (see Fig.\,\ref{fig:im70}) by surveying the {\it Spitzer}-GLIMPSE,  
 {\it Spitzer}-MIPSGAL, and {\it Herschel}-HiGAL archives from the IRSA data base\footnote{\url{http://irsa.ipac.caltech.edu/frontpage/}}. 
 We then performed the YSO SED fitting for the six sources 
 following the procedures as described in \citet{liu16}. The resulting fitting plots are presented in Fig.\,\ref{fig:YSOfit}.
 Following the descriptions by \citet{yua14}, we calculated several key parameters, which are tabulated in Table\,\ref{tbl:YSO}.
 As shown in Fig.\,\ref{fig:YSOfit}, the fitting performances are not being perfect. This can be attributed to at least the following three aspects. Firstly, for the {\it Herschel} photometry data, poor resolutions at wavelengths of $\geq160$\,\um\ make it difficult to accurately measure the photometry fluxes that really represent radiation from YSOs. Secondly, the YSO models in 
 the SED fitting tool can not interpret the photometry fluxes at longer wavelengths very well (Robitaille, private communication). Thirdly, as a result of the previous two problems,
the more the data points included, the larger the goodness of the SED fitting. As defined in \citet{liu16}, the goodness ($\chi2/N_{\rm data}$ where $N_{\rm data}$ is the number of samples 
included in the fitting) of the SED fitting for the six bright sources ranges from 5 to 38 (see Table\,\ref{tbl:YSO}). Despite the goodness of the fitting not perfect, some of the resulting 
fitting parameters can still be used to get some hints from the statistical point of view. 

According to the scheme of YSO classification by \citet{rob06}, all six sources can be classified as
Stage\,0/I YSO candidates (see Table\,\ref{tbl:YSO}). Their spatial associations with the dense cores are commented in
Table\,\ref{tbl:YSO}. Such associations are well supportive of star formation ongoing in some of the dense cores. 
In addition, these YSO candidates have masses ranging from 0.7 to 2.3\,\msun. This mass range
demonstrates that there are already histories of low-mass star formation in some of nine dense cores. All in all, the mass-size distributions of the nine 
dense cores, and the low-mass YSOs associated with some cores collectively suggest that both filaments could be sites of ongoing low-mass star formation.

\subsection{Comparison with other two filaments}
\label{sec:comparison}
\begin{figure*}
\centering
\includegraphics[width=6.9 in]{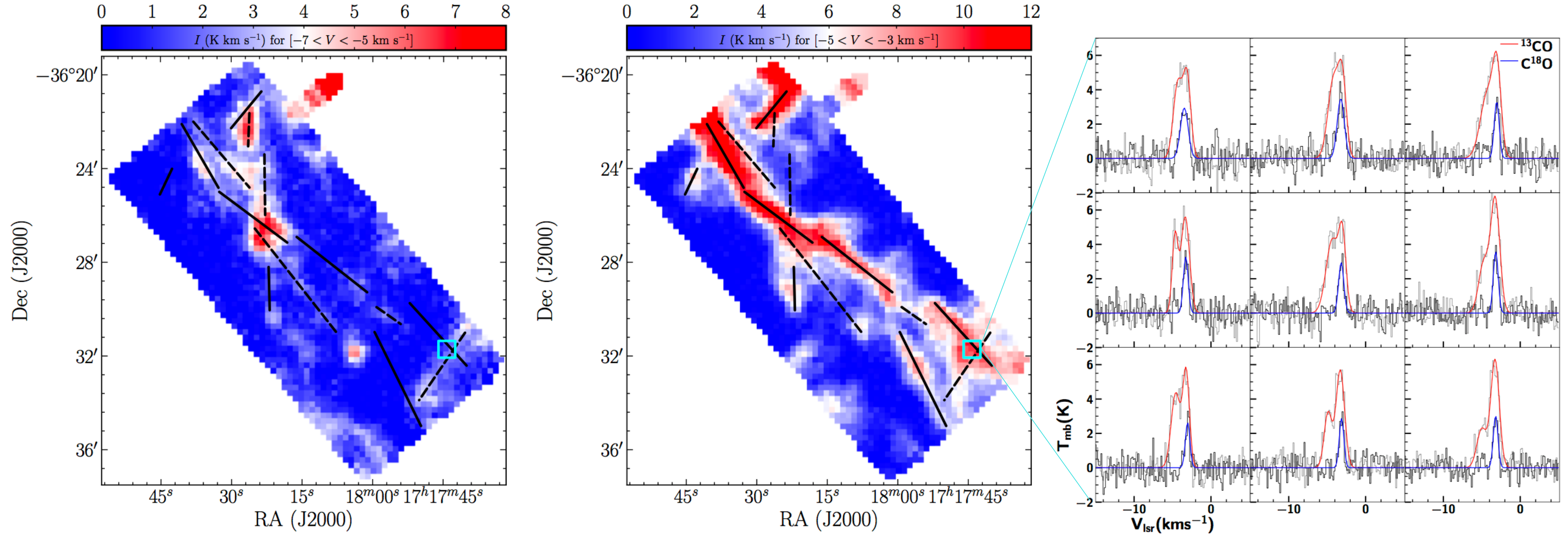}
\caption{Velocity-integrated intensity of \thco~(2-1) over the two velocity ranges, $-7$~to~$-5$\,\vel\ in ({\it Left}) and $-5$~to~$-3$\,\vel\ in ({\it Middle}).  
 In both panels, the dashed lines delineate the filament branches appearing at $-7$~to~$-5$\,\vel\ and
the solid lines represent the branches for $-5$~to~$-3$\,\vel. ({\it Right:}) spectra of \thco\ and \etco~(2-1) for a selected region. 
The selected region is marked with the cyan box in the {\it Left} and {\it Middle}  panels. Black lines are the observed spectra and the color lines are the
Gaussian-fitting curves (red for \thco\ and blue for \etco). This panel shows the two velocity components in 
\thco~(2-1), one of which matches well with the single component revealed by relatively optically thin \etco~(2-1).
}
\label{fig:filabranch}
\end{figure*}

\filname\ appears basically straight and isolated, with similarities to both  Musca and Taurus B211/3. Moreover, \filname\ is surrounded by 
thin perpendicular striations, especially in the \filAname\ part (see Fig.\,\ref{fig:im70}, \ref{fig:imHer}), which
resembles the dust emission morphologies in Musca, and Taurus B211/3 \citep[e.g.,][]{pal13,cox16}. This resemblance indicates that
\filname\ might have a similar formation mechanism to Musca, and Taurus B211/3. 

Based on magnetically-aligned striations observed in Taurus B211/3, 
and Musca, \citet{hen13} and \citet{cox16} proposed a possible filament formation mechanism. That is, the filament grows in mass by accreting background material channeled
by magnetic fields, leading to accretion-driven MHD turbulence within the main filament structure \citep{hen13,cox16}. The accretion transfers 
gravitational energy to the system, which sequentially turns into turbulent kinetic energy \citep{cox16}. During the accretion, 
a supercritical filament would evolve into a more complicated network of intertwined fibers as the internal velocity dispersion of the filament
increases, which could provide us with a hint on the evolution of the filaments. Due to the absence of complicated intertwined 
velocity-coherent fibers in Musca \citep{kai16,har16} as observed in Taurus B211/3 \citep{har13,taf15}, \citet{cox16} suggested that Musca 
is at an earlier stage of evolution than Taurus B211/3 filament, which is further supported by a very low present-day star-formation 
efficiency of $\ll 1 \%$ derived from a candidate young low-mass T Tauri star located at the most northeast end of the Musca filament.

Figure\,\ref{fig:filabranch} shows the velocity integrated intensity map of the cloud \filname\ over the two velocity ranges 
$-7$~to~$-5$\,\vel, and $-5$~to~$-3$\,\vel. \filname\ does not seem to have a complicated network of velocity-coherent intertwined 
fibers within the main filament structure, suggestive of the cloud \filname\ {\it being} likely less evolved than Taurus B211/3.
Note that our current molecular line observations with the resolution $28\arcsec$ probably do not resolve the intrinsically complicated network 
of velocity-coherent fibers in \filname. Therefore, future dense gas observations with higher resolutions would be
helpful for dissecting the internal velocity structures within the main filament structure of the cloud \filname.
Regardless of the velocity-coherent structures, the presence of different stages of YSOs in the two clouds where \filname\ has Stage 0/I YSOs only but the Taurus B211/3 cloud 
has YSO candidates evolved up to T\,Tauri stars, also supports that \filname\ could be in an intermediate evolutionary state between Musca and Taurus-B211/3.

\section{Conclusions}

We have investigated the column density profiles of the two filaments \filAname\ and \filBname, the gravitational stability of the dense cores on the filaments, possible
fragmentation process, and 
star formation in the two filaments, combing {\it Herschel} data with our molecular line observations by APEX. The main findings are summarized here:

\begin{itemize}
\item The filamentary cloud \filname\ appears rather straight and basically isolated, 
composed of two discontinuous filaments, \filAname\ and \filBname. \filAname~(S) has a length of $\sim5.9$\,pc~($\sim2.3$\,pc), 
a total mass of $\sim 810$\,\msun~($\sim 110$\,\msun),
and a mean temperature of  $\sim 18.2$\,K~($\sim 17.7$\,K). Both filaments have a similar mean column density of $\sim 8.2\times10^{22}$\,\cmcm.

\item The mean column density profiles of \filAname, and \filBname\ are described in a Plummer-like function with a power law index of $p \sim 3$. 
This index is only a first-order approximation. Dust continuum observations with higher spatial resolutions deserve to reveal the
detailed density profiles of both filaments.

\item Nine dense cores are identified with $M_{\rm core} \geq M_{\rm BE}$. That is, the cores are gravitationally bound and have 
the potential to further collapse to form protostars.

\item The separation ($\sim 0.9$\,pc) among nine cores  and the average line mass ($\sim 101$\,\lmsun) of the whole cloud 
appear to follow the predictions of the  ``sausage'' instability theory.  This suggests that \filname\ could have undergone radial collapse and fragmentation.

\item The mass-size relation of the nine cores, and the appearance of six low-mass YSOs associated with the cores together provide strong evidence 
 that \filname\ is a site of ongoing low-mass star formation.

\item The presence of young protostars (i.e., Stage\,0/I)
indicates that the G350.5 filament may be in an evolutionary state between Musca, with no protostars, and Taurus B211/3, with 
evolved young stars (i.e., class\,II YSOs).

The filamentary cloud \filname\ has not been well-studied until now. Its rather straight and isolated feature may be very helpful for dissecting the filament
and understanding the nature of isolated and straight filament formation.
Our forthcoming paper (Liu et al., 2018, in prep.), will be devoted to further investigation of kinematics of this system, 
which will allow us to understand the connection between those two filaments and core formation therein.

\end{itemize}

\medskip
\noindent{\textbf{Acknowledgments}}\\
  This work was in part sponsored by the Chinese Academy of Sciences (CAS), through a grant to
the CAS South America Center for Astronomy (CASSACA) in Santiago, Chile.
  AMS acknowledges funding from Fondecyt regular (project code 1180350),  the ''Concurso Proyectos 
  Internacionales de Investigaci\'on, Convocatoria 2015'' (project code PII20150171), and the BASAL 
  Centro de Astrof\'isica y Tecnolog\'ias Afines (CATA) PFB-06/2007. J. Yuan is supported by the National 
  Natural Science Foundation of China through grants 11503035, 11573036.We thank the anonymous referee 
  for the comments that much improved the quality of this paper.

   SPIRE has been developed by a consortium of institutes
led by Cardiff Univ. (UK) with Univ. Lethbridge (Canada); NAOC (China);
CEA, LAM (France); IFSI, Univ. Padua (Italy); IAC (Spain); Stockholm
Observatory (Sweden); Imperial College London, RAL, UCL-MSSL, UKATC,
Univ. Sussex (UK); Caltech, JPL, NHSC, Univ. Colorado (USA). This development
has been supported by national funding agencies: CSA (Canada);
NAOC (China); CEA, CNES, CNRS (France); ASI (Italy); MCINN (Spain);
SNSB (Sweden); STFC (UK); and NASA (USA). PACS has been developed
by a consortium of institutes led by MPE (Germany) with UVIE (Austria); KU
Leuven, CSL, IMEC (Belgium); CEA, LAM(France); MPIA (Germany); INAFIFSI/
OAA/OAP/OAT, LENS, SISSA (Italy); IAC (Spain). This development
has been supported by the funding agencies BMVIT (Austria), ESA-PRODEX
(Belgium), CEA/CNES (France), DLR (Germany), ASI/INAF (Italy), and
CICYT/MCYT (Spain). We have used the NASA/IPAC Infrared Science Archive
to obtain data products from the \emph{Spitzer}-GLIMPSE, and \emph{Spitzer}-MIPSGAL
surveys.

This research made use of astrodendro, a Python package to compute dendrograms of Astronomical data (\url{http://www.dendrograms.org/}). 
Also, this research made use of Astropy, a community-developed core Python package for Astronomy (Astropy Collaboration, 2018).

%\vspace{-5mm}
\bibliographystyle{mnras}
\bibliography{../paper}
% 
% \clearpage
 \appendix
\section{Continuum emission at 160, 250, 350, and 500\,\um}
\begin{figure*}
\centering
\includegraphics[width=0.26\textwidth]{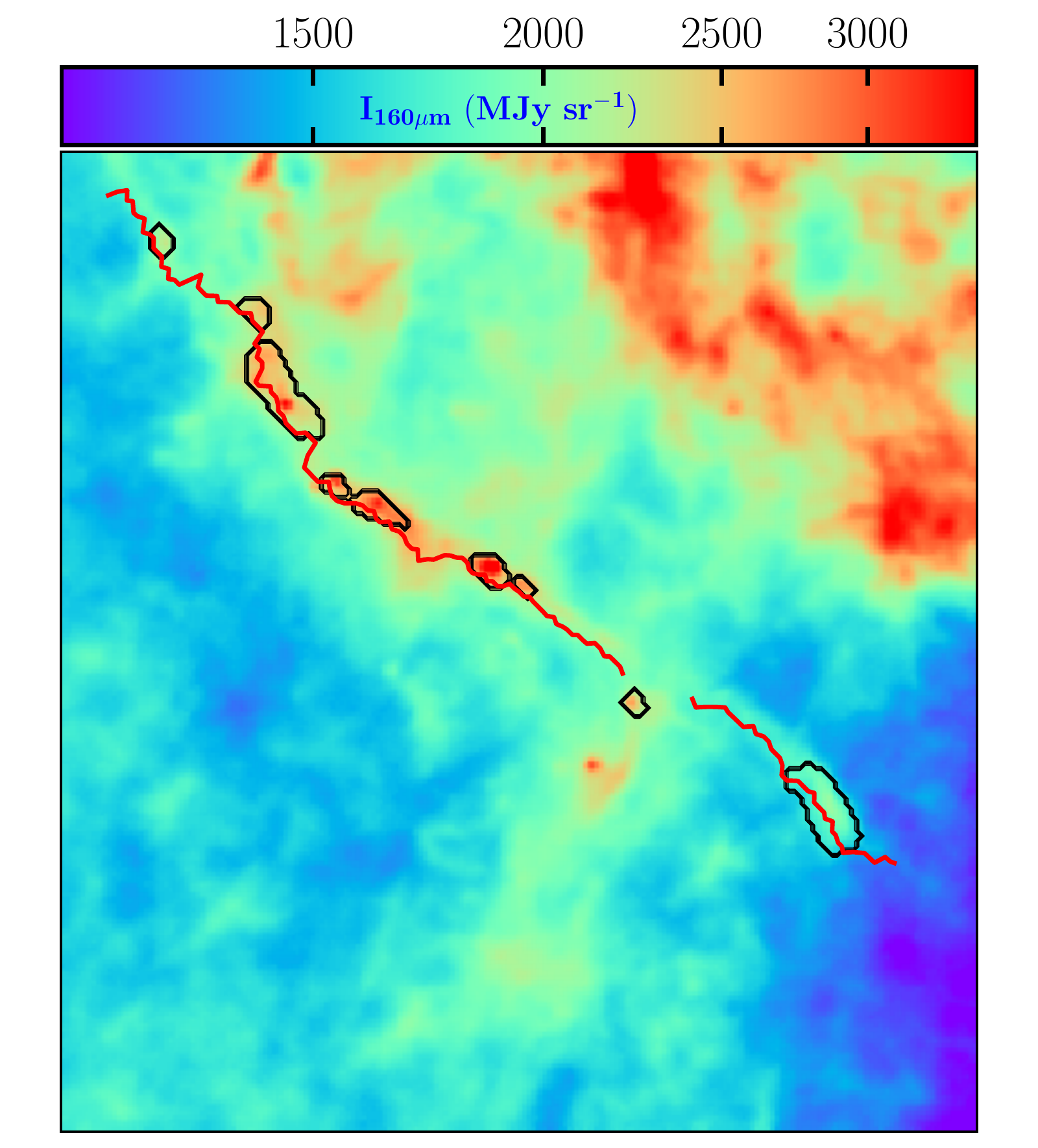}
\hspace{-0.39cm}
\includegraphics[width=0.26\textwidth]{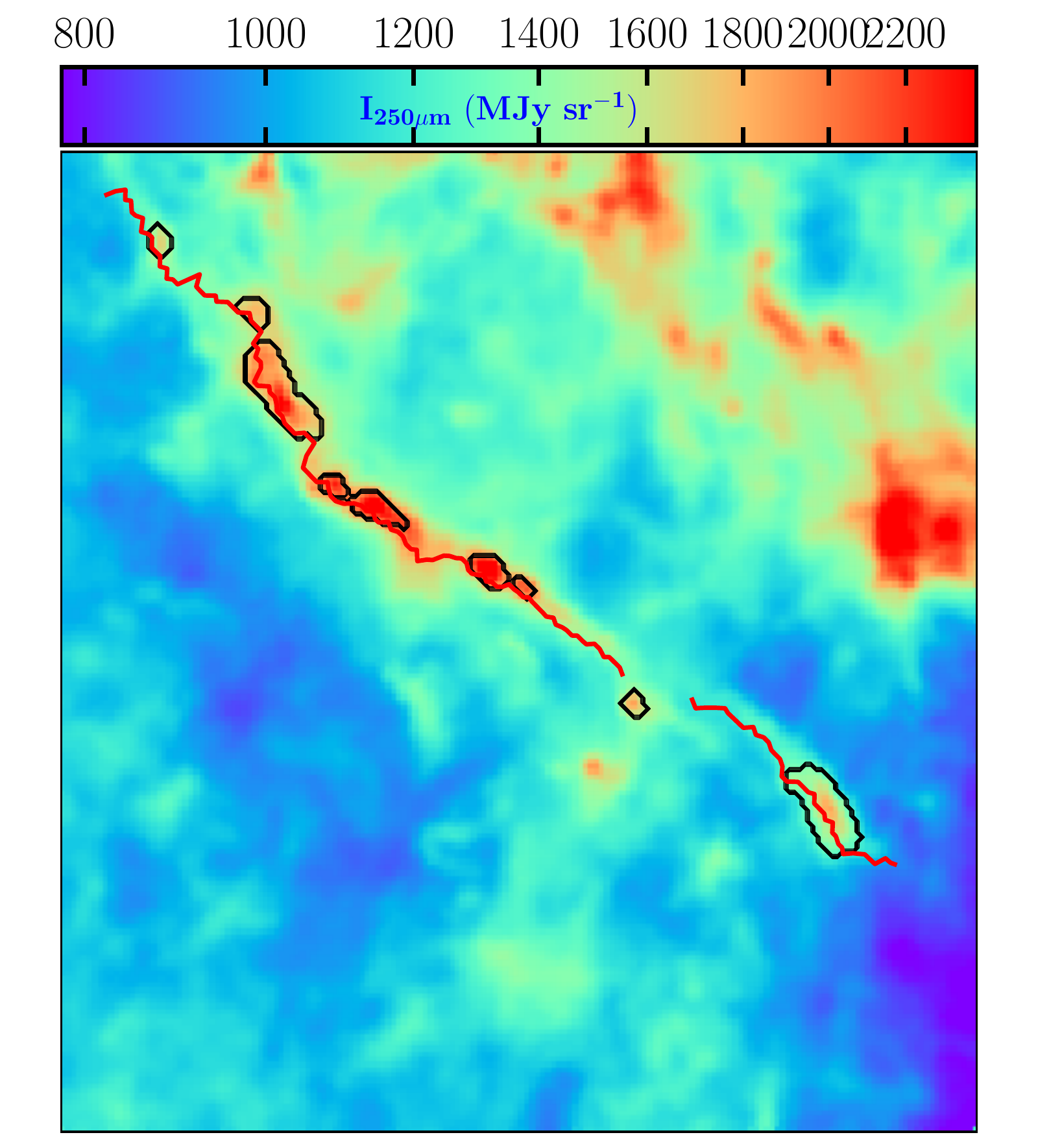}
\hspace{-0.39cm}
\includegraphics[width=0.26\textwidth]{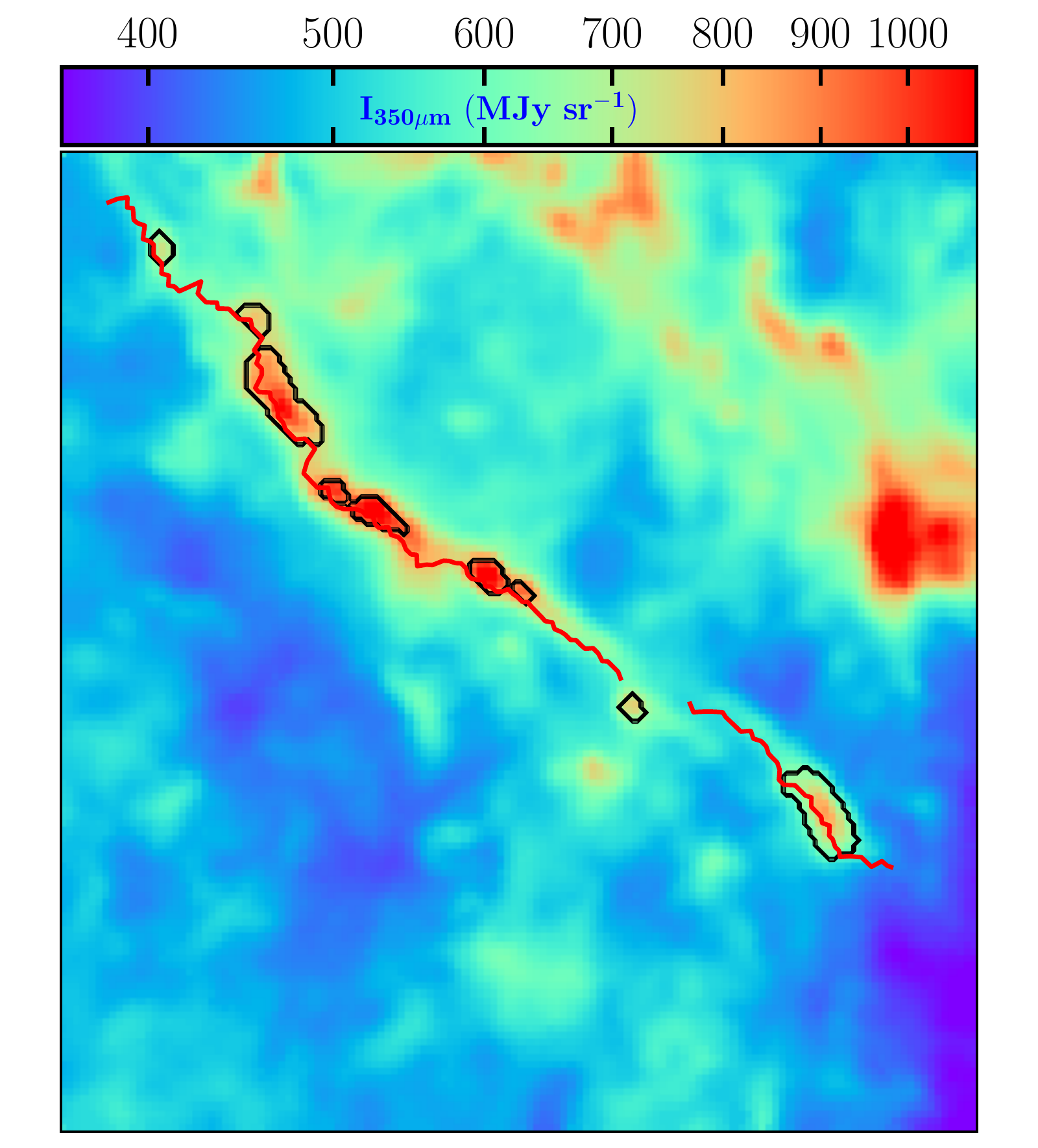}
\hspace{-0.39cm}
\includegraphics[width=0.26\textwidth]{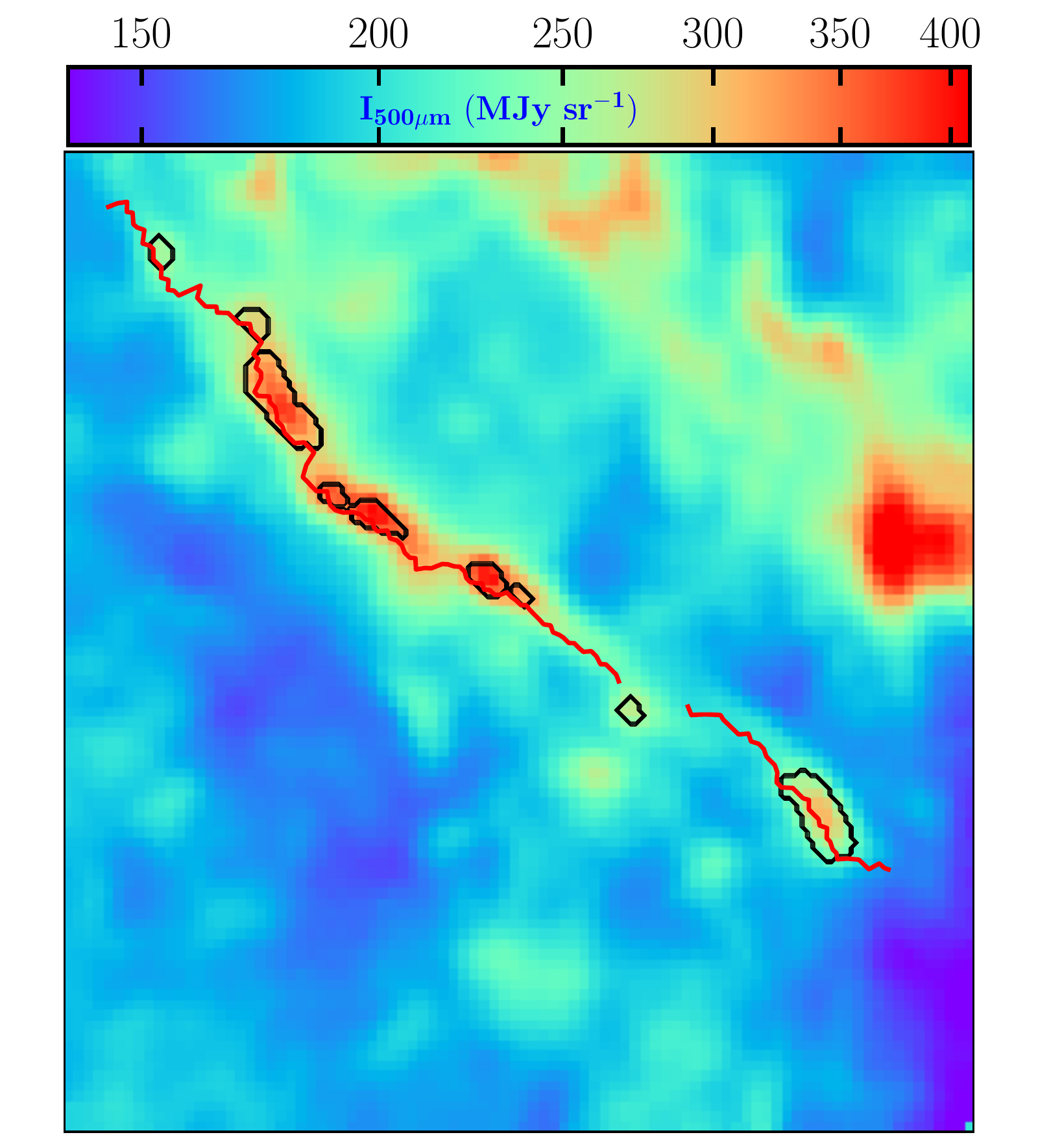}
\caption{ Same as Fig.\,\ref{fig:im70} but for \her\ 160, 250, 350, 500\,\um\ from left to right.
The main structure of G350.5 appears to be surrounded by thin perpendicular striations. This scenario
seems to be clearer at both 160 and 250\,\um\ than at both 350 and 500\,\um, which
could be a consequence of lower spatial-resolution at longer wavelengths. 
}
\label{fig:imHer}
\end{figure*}

\bsp
\label{lastpage}
\end{document}